\documentclass[useAMS,usenatbib]{mn2e}
\usepackage{graphicx}
\usepackage{scalefnt}
\usepackage[]{booktabs}
\usepackage{colortbl}
\usepackage[table]{xcolor}
\usepackage{helvet}
\usepackage{multimedia}
\usepackage{long table}

\title[Dynamical evolution of 
V-type photometric candidates in the central main belt]
{Dynamical evolution of V-type asteroids in the central main belt}
\author[V. Carruba, M. E.Huaman, R. C. Domingos, C. R. Dos Santos, 
and D. Souami]{ V. Carruba$^{1}$\thanks{E-mail: vcarruba@feg.unesp.br}, M. E. 
Huaman$^{1}$, R. C. Domingos$^{2,1}$, C. R. Dos Santos$^{1}$, and D. 
Souami$^{3,4,5}$\\
$^{1}$UNESP, Univ. Estadual Paulista, Grupo de din\^{a}mica Orbital e
  Planetologia, Guaratinguet\'{a}, SP, 12516-410, Brazil \\
$^{2}$INPE, Instituto Nacional de Pesquisas Espaciais,
  São José dos Campos, SP, 12227-010, Brazil \\
$^{3}$NAXYS, Namur Center for Complex Systems, Department of Mathematics, 
University of Namur, 5000 Namur, Belgium\\
$^{4}$UPMC, Universit\'{e} Pierre et Marie Curie, 4 Place Jussieu, 75005, 
Paris, France\\
$^{5}$SYRTE, Observatoire de Paris, Syst\`{e}mes de R\'{e}f\'{e}rence 
Temps Espace, CNRS/UMR 8630, UPMC, Paris, France;
}

\begin{document}

\date{Accepted 2014 January 24.  Received 2014 January 23; 
in original form 2013 December 9}

\pagerange{\pageref{firstpage}--\pageref{lastpage}} \pubyear{2010}

\maketitle

\label{firstpage}

\begin{abstract}
V-type asteroids are associated with basaltic composition, and are 
supposed to be fragments of crust of differentiated objects. Most 
V-type asteroids in the main belt are found in the inner main belt, and 
are either current members of the Vesta dynamical family (Vestoids), or past 
members that drifted away.  However, several V-type photometric 
candidates have been recently identified in the central and 
outer main belt.

The origin of this large population of V-type objects is 
not well understood.  Since it seems unlikely that Vestoids
crossing the 3J:-1A mean-motion resonance with Jupiter could account
for the whole population of V-type asteroids in the central and outer
main belt, origin from local sources, such as the parent bodies 
of the Eunomia, and of the Merxia and Agnia asteroid families, has been 
proposed as an alternative mechanism.

In this work we investigated the dynamical evolution of the V-type
photometric candidates in the central main belt, under the effect 
of gravitational and non-gravitational forces. Our results show that 
dynamical evolution from the parent bodies of the Eunomia and 
Merxia/Agnia families on timescales of 2 Byr or more could be responsible for
the current orbital location of most of the low-inclined V-type asteroids.

\end{abstract}

\begin{keywords}
Minor planets, asteroids: general -- Minor planets, asteroids: classes: 
V-types -- celestial mechanics.  
\end{keywords}
%
%________________________________________________________________

\section{Introduction}
\label{sec: intro}

The V-type asteroids, which are characterized by 1 and 2 $\mu m$
absorption bands in the infrared spectrum, are associated with
basaltic composition, and are supposed to be fragments of the crust of 
differentiated objects.  Most of the V-type asteroids in the main 
belt are found in the inner main belt (defined as the 
region in semi-major axis between the 4J:-1A and 3J:-1A mean-motion 
resonances with Jupiter), and are either members of
the dynamical Vesta family (Vestoids), or are thought to be
past members that dynamically migrated beyond the limits of the 
current Vesta family.  Recently, two new basaltic asteroids, (1459) Magnya
(Lazzaro et al. 2000), and (10537) (1991 RY16) (Moskovitz et
al. 2008b), were identified in the outer main belt (the region between 
the 5J:-2A and 2J:-1A mean-motion resonances),
a region too far away from Vesta to possibly be dynamically 
connected with this asteroid.  Roig and Gil-Hutton (2006) used 
Sloan Digital Sky Survey-Moving Object Catalog data, fourth release (SDSS-MOC4) 
multi-band photometry to 
infer the asteroid taxonomy, and identified several V-type
photometric candidates in the central (defined as the 
region between the 3J:-1A and 5J:-2A mean-motion resonances)
and outer main belt.
Most recently, Carvano et al. (2010) identified hundreds of photometric
candidates in the central and outer main belt.  Roig et al. (2008)
discussed the possibility that some of these asteroids could be
former Vestoids that crossed the 3J:-1A mean-motion resonance with Jupiter,
but this mechanism alone seems to be unlikely to have produced the 
whole observed population of V-type candidates, especially those
in the outer part of the central main belt.

Alternative production mechanisms have therefore been suggested to 
explain the presence of these objects.  Carruba et al. (2007a) suggested
that three V-type asteroids in the central main belt, (21238), (40621)
and (66905) could have originated as fragments of the crust of
the parent body of the Eunomia family, that, according to Nathues
et al. (2005) might have been a differentiated or partially differentiated
body.  This mechanism may have originated some of the basaltic material
in the central main belt.  However, since the work of Carvano et al. (2010)
identified more than one hundred V-type photometric candidates in the central
main belt, and since some of these objects have very different values
of inclination with respect to (15) Eunomia, yet another mechanism
to produce basaltic material may have been at work.

Luckily, other possible sources of basaltic material were identified
in the central main belt by other authors.
Observations of (808) Merxia and (847) Agnia by Sunshine et al. (2004)
showed that the spectra of both these objects are compatible 
with an S-type asteroid with both low and high calcium forms 
of pyroxene on the surface, along with less than 20\% olivine. 
The high-calcium form of pyroxene forms 40\% or more of the total 
pyroxene present, indicating a history of igneous rock depopsits. 
This suggested that the asteroids underwent differentiation by melting, 
creating a surface of basalt rock.

The members of these families, including their namesakes, 
according to Sunshine et al. (2004), 
most likely formed from the breakup of a basalt object, 
which in turn was spawned from a larger parent body that had 
previously undergone igneous differentiation.
Since the Merxia and Agnia families have lower values of inclinations 
than (15) Eunomia, could it be possible that some of the V-type
photometric candidates currently at low-$i$ may have originated
from the parent body of these two families?

In this work we will investigate possible paths of diffusion from
the current orbital location of the photometric candidates 
in the central main belt, in order to infer possible clues on 
their origin.  Since Moskovitz et al. (2008a) suggested that 
it was not likely to have had hundreds of differentiated
bodies in the primordial main belt, but that the number was more
likely of the order of dozens, we will try to assess how much 
a model with two sources in the central main belt, the parent 
bodies of the Eunomia and that of the Merxia/Agnia families, may succeed or not
in reproducing the current distribution of V-type photometric candidates.

This paper is so divided: in Sect.~\ref{sec: taxonomy-sdss} we discussed
the current knowledge on V-type photometric candidates in the central 
main belt, obtained in previous works (Roig and Gil-hutton 2006, Carvano et 
al. 2010, De Sanctis et al. 2011) and by our group using the approach 
of DeMeo and Carry (2013).  Five different orbital regions characterized
by the presence of basaltic material are introduced in this section.
In Sect.~\ref{sec: V-type_groups} we discuss the groups identified
in Sect.~\ref{sec: taxonomy-sdss} in terms of ejection velocities
with respect to Eunomia, diameters, geometrical albedos, and masses.
In Sect.~\ref{sec: dyn_maps} we study the local web of mean-motion 
and secular resonances using dynamical maps of synthetic proper
elements.  In Sect.~\ref{sec: yarko} we analyze how dynamical
evolution caused by the Yarkovsky effect may have caused the current
orbital distribution of V-type asteroids, and 
in Sect.~\ref{sec: close_encounters} we study the long-term effects of close 
ecnounters with massive asteroids in the central main belt.
Finally, in Sect.~\ref{sec: concl} we present our conclusions.

%%%%%%%%%%%%%%%%%%%%%%%%%%%%%%%%%%%%%%%%%%%%%%%%%%%%%%%%%%%%
\section{Identification: spectral Taxonomy and SDSS-MOC4 data}
\label{sec: taxonomy-sdss}

Carvano et al. (2010) studied the taxonomy of asteroids observed 
by the Sloan Digital Sky 
Survey-Moving Object Catalog data, fourth release (SDSS-MOC4 
hereafter, Ivezic et al. 2001).  This survey provided 
a sample two order of magnitude larger than any available in any 
current spectroscopic catalogs (about 60000 numbered objects).  
The authors identified 130 V-type photometric candidates in 
the central main belt, including
QV, SV objects, and other dubious cases, in the central
main belt, based on a taxonomical scheme in the plane of $(a^{*}, i-z)$
colors, and posted their results on the Planetary Data System, 
available at http://sbn.psi.edu/pds/resource/sdsstax.html, and accessed 
on September $10^{th}$ 2013.

DeMeo and Carry (2013) recently introduced
a new classification method, based on the Bus-DeMeo taxonomic
system, that employs SDSS-MOC4 gri slope and $z^{'} -i^{'}$ colors. In
that article the authors used the photometric data obtained in the
five filters $u^{'} , g^{'} , r^{'} , i^{'}$, and $z^{'}$, from 0.3 to 1.0 $μm$, 
to obtain values of $z^{'} -i^{'}$ colors and spectral slopes over 
the $g^{'}, r^{'}$, and $i^{'}$ reflectance values.  Values of 
$z^{'} - i^{'}$ colors 
and spectral slopes were then used to assess if an asteroid belonged
to a given spectral class, whose boundary values were determined
in this domain by the authors.  For overlapping classes, the taxonomy 
is assigned in the last class in 
which the object resides, in the following order: C-, B-, S-, L-, X-,
D-, K-, Q-, V-, and A-types. Since some asteroids had multiple 
observations in the SDSS-MOC4 database, we adopted the DeMeo and 
Carry (2013) criteria for classifications in these cases: in case of
conflicts, the class with the majority number of classifications
is assigned.

With respect to the work of Carvano, we identified 
six new photometric candidates
in the central main belt, and eleven in the outer main belt.  
Table~\ref{table: aster_V}
displays our new results.  All V-type photometric candidates
identified by Carvano et al. (2010) are also listed 
in Table~\ref{table: V_carvano}.
We report the asteroid identification,
if it is a confirmed V-type object, the asteroid proper 
elements $a,e,$ and $sin(i)$,
the absolute magnitude ($H$), the diameter ($D$), and geometric
albedo ($p_V$), according to the WISE mission, when available.   The asteroid 
(177904) (2005 SV5) was already identified as a possible V-type photometric
candidate in Carruba et al. (2013b).  Being this a paper centered on 
the dynamical evolution of V-type candidates in the central main belt only, 
asteroids in the outer main belt are only given as a reference to the 
observing comunity,

%%%%%%%%%%%%%%%%%%%%%%%%%%%%%%%%%%%%%%%%%%%%%%%%%%%%%%%%%%%%
\begin{table*}
\begin{center}
\caption{{\bf List of new V-type photometric candidates 
in the central and outer Main-belt.}}
\label{table: aster_V}
\vspace{0.02cm}
\begin{tabular}{r l c c c c c c}        % centered columns (8 columns)
\hline\toprule 
Asteroid Id. & Spectral type  & $a$ & $e$ & $sin(i)$  & $H$ & $D$ (km) & $p_{v}$ \\
\midrule %\hline
%%%%%%%%%%%%%%%%%%%%%%%%%%%%%%%%%%%%%%%%%%%%%%%%%%%%%%%555
    7472&  V?   &    3.010245&       0.138886&       0.156293&   12.08&10.008 & 0.2795\\
   11465&  V?   &    3.102607&       0.014941&       0.348060&   12.87&12.556 & 0.0775\\
   18802&  V?   &    2.687340&       0.143116&       0.215531&   13.82& 3.940 & 0.3134\\
   35567&  V    &    2.567894&       0.089490&       0.196503&   14.58& 2.381 & 0.4504\\
   55270&  V    &    2.908541&       0.125003&       0.164700&   14.02&&\\
   57322&  V?   &    2.584340&       0.164743&       0.130332&   14.86&&\\
   88084&  V?   &    2.545195&       0.152695&       0.088088&   15.30&&\\
   91159&  V    &    3.194682&       0.273886&       0.277364&   14.28&&\\
   92182&  V?   &    3.181359&       0.169964&       0.266076&   13.70&&\\
   93981&  V    &    2.662526&       0.019289&       0.226328&   13.85& 3.376 & 0.4268\\
  177904&  V    &    3.158618&       0.153963&       0.073138&   15.30& 5.762 & 0.0403\\
  208324&  V    &    3.130781&       0.270397&       0.287020&   14.91&&\\
  217953&  V?   &    2.546619&       0.162840&       0.176317&   16.37&&\\ 
\bottomrule%\hline
\end{tabular}
\end{center}
\end{table*}

We selected all asteroids in the Carvano et al. (2010) database, 
available in the  the Planetary Data System, 
and in the HORIZONS system of the Jet Propulsion laboratory
~\footnote{The Database was accessed on September $13^{th}$~2013.}, 
that were V-type photometric candidates (including SV and QV types),
had proper elements as reported by the AstDyS site 
http://hamilton.dm.unipi.it/cgi-bin/astdys/astibo, 
(accessed on September $15^{th}$, 2013, Kne\v{z}evi\'{c} and Milani 2003), 
and had orbits in the central main belt, defined
as $a_{31} < a < a_{52}$, where $a_{31}$ and $a_{52}$ are the semi-major
axis of the center of the 3J:-1A and 5J:-2A mean motion resonances
with Jupiter, respectively.  Overall, including the new photometric
candidates identified in this work, we encountered 255 V-type
candidates, of which 127 are pure V-types (excluding SV, QV, and
other dubious cases), and objects, (21238), 
(40521), and (66905) are confirmed V-type asteroids whose spectra 
were obtained in De Sanctis et al. (2011).
Our results are shown in Fig.~\ref{fig: V-type_central}, where
we display a proper $(a,e)$ (panel A) and $(a,sin(i))$ projections
of the proper elements of all candidates (blue dots), of pure V-type
candidates (green full dots) and of the confirmed V-type asteroids
(red full dots).  Vertical lines identify the location of the main 
local mean-motion resonances, blue lines show the position of the 
center of secular resonances, computed using the analytical theory 
of Milani and Kne\v{z}evi\'{c} (1994)
to compute the proper frequencies $g$ and $s$ for the grid of $(a,e)$
and $(a,\sin(i))$ values shown in Fig.~\ref{fig: V-type_central} and the
values of angles and eccentricity of (480) Hansa, the asteroid with the 
largest family in the highly inclined region.
In panel A, we also display the lines for which $q=Q_{Mars}$, and $q=q_{Mars}$,
where close encounters with terrestrial planets start to be possible.

\begin{figure*}

  \centering
  \begin{minipage}[c]{0.5\textwidth}
    \centering \includegraphics[width=3.5in]{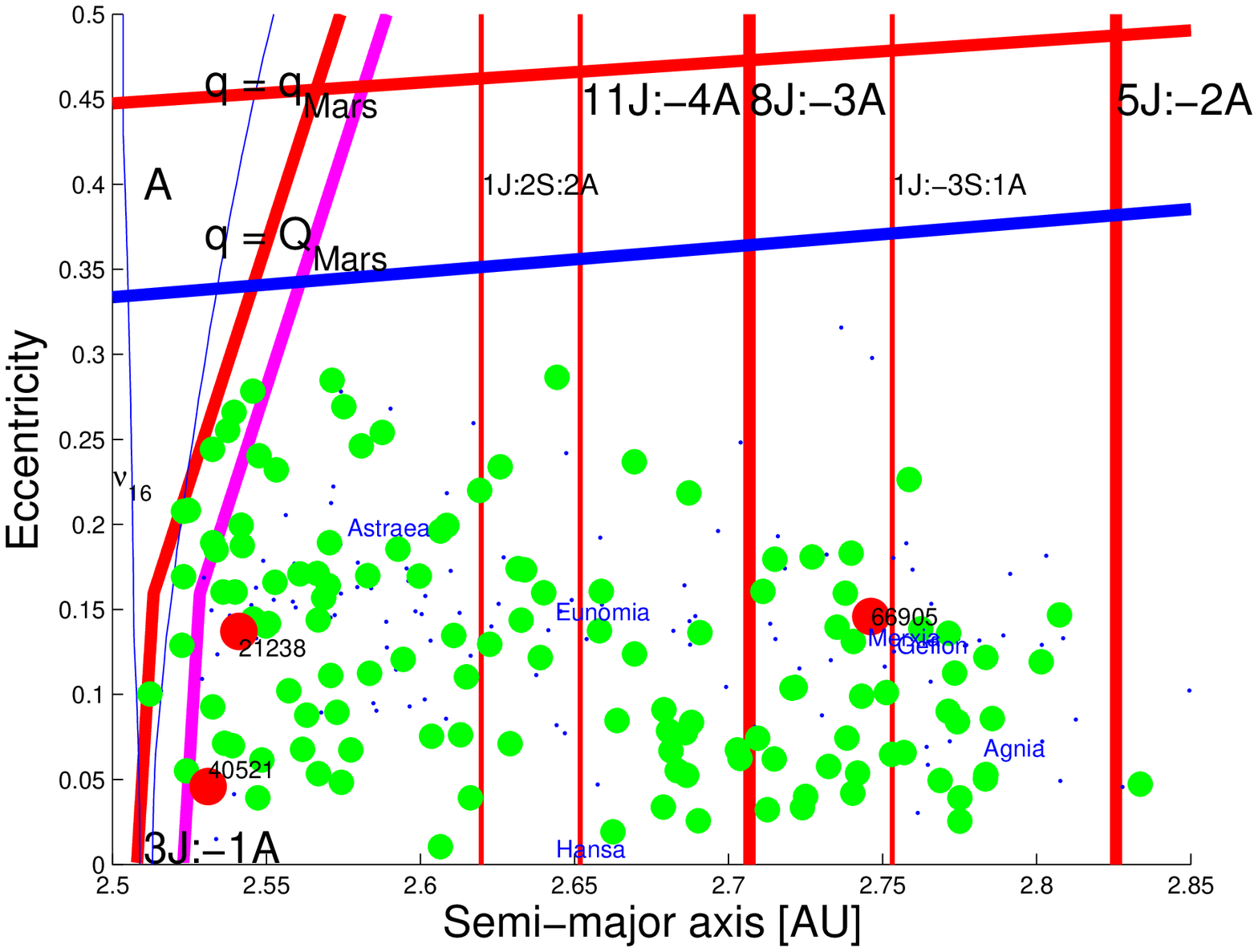}
  \end{minipage}%
  \begin{minipage}[c]{0.5\textwidth}
    \centering \includegraphics[width=3.5in]{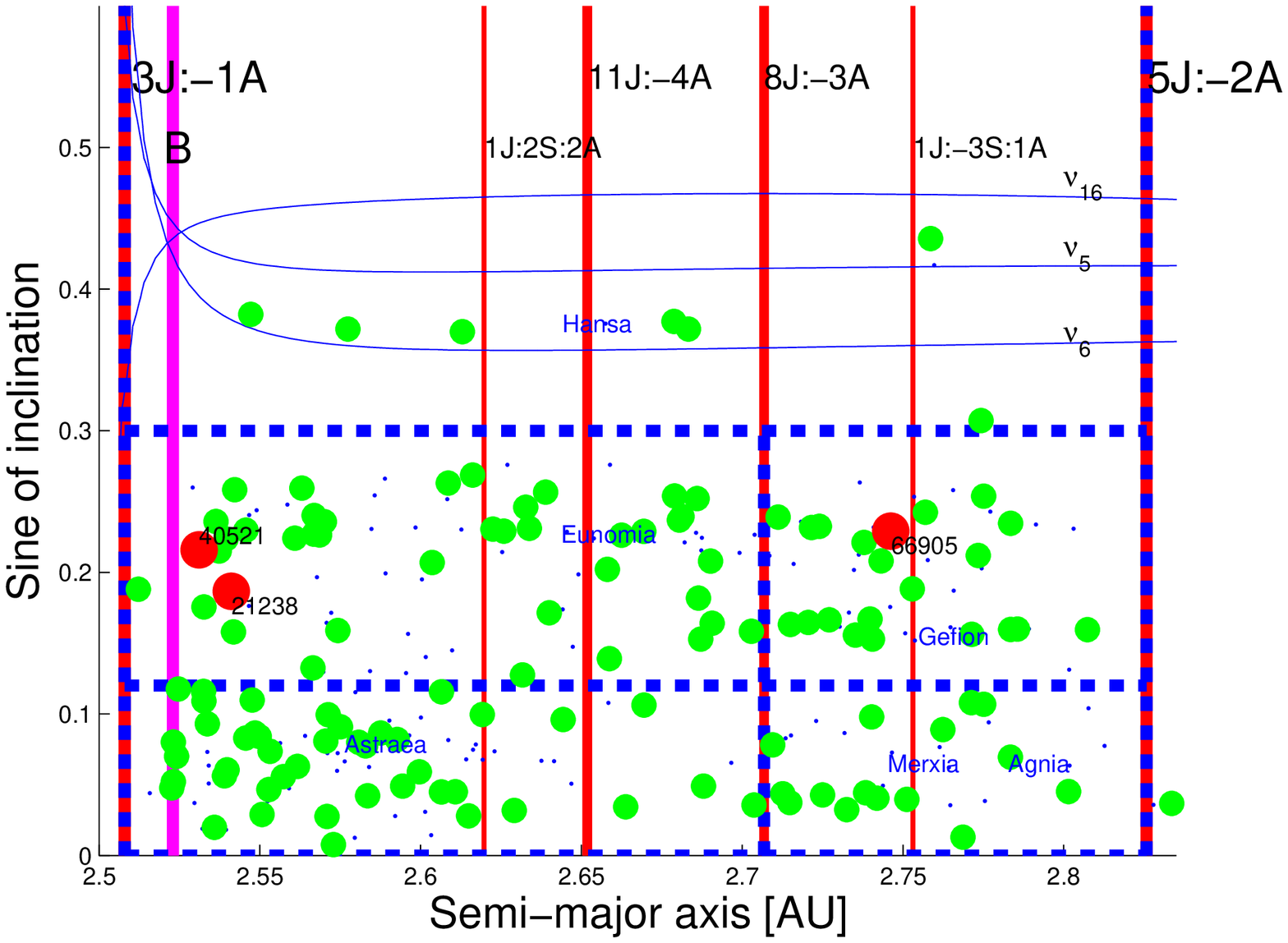}
  \end{minipage}

\caption{Panel A: Semi-major axis versus eccentricity of the V-type candidates
	identified using SDSS-MOC4 in the central main belt.
Blue dots identified the orbital location of all photometric
candidates, green full dots are associated with ``pure'' V-type 
candidates, and confirmed V-types are shown as red full dots.
Vertical red lines display the location of mean-motion resonances,
and blue lines show the center of secular resonances (see text for
details).  The magenta line is associated with the unstable region
(on timescales of 100 Myr) near the 3J:-1A mean-motion resonance, 
as identified by Guillens {\em et al.} (2002).  Finally,
we also display the lines for which $q=Q_{Mars}$, and $q=q_{Mars}$,
where close encounters with terrestrial planets start to be possible.
Panel B: Semi-major axis versus sine of proper inclination of the same 
V-type candidates.  Blue dotted lines identify the boundaries
of the regions discussed in the text.}
\label{fig:  V-type_central}
\end{figure*}

One may notice six regions of concentrations of V-type candidates in the
$(a,sin(i))$ plane:  one region around the Eunomia family, in the inner
central main belt ($a_{31} < a < a_{83}$), that includes the three
confirmed V-type asteroids discussed in Carruba et al. (2009), at 
intermediate values of $sin(i)$ ($0.12 < sin(i) < 0.3$).
A concentration of asteroids near the Astraea family, in the inner
central main belt, at low inclinations ($sin(i) < 0.12$). 
Eight asteroids in the highly inclined region ($sin(i) > 0.3$),
mostly found in the Hansa family region. Finally, three alignments
of asteroids (possibly four, if we include less than ``pure'' V-type
candidates), in the outer central main belt ($a_{83} < a < a_{52}$), 
in the regions of the Merxia-Agnia, Gefion, and (1995 SU37) asteroid families,
with values of $sin(i)$ lower than 0.12 (Lavrov), $0.12 < sin(i) < 0.19$
(Gefion), and $sin(i)> 0.19$, respectively.   
We define these regions as Hansa, Eunomia, Astraea, Merxia, and 
Gefion~\footnote{We consider the two alignments
in the regions of the Gefion and (1995 SU37) asteroid families
to be part of the same ``Gefion'' region.}, 
and assign each V-type candidate to its own
region, whose boundaries are shown in Fig.~\ref{fig:  V-type_central}, 
panel B.  Results are also listed in Table~\ref{table: V_carvano}.
We also checked how many of the known ``pure'' V-type candidates are
members of known asteroid families, according to the AstDyS site.
We found 22 asteroids belonging to a family, most of which (six, 
27.27\% of the total) members of the Eunomia family, followed by four
objects (18.18\%) in the Hansa family.  Of the remaining asteroids
belonging to identified families,
two were in the Juno family, one in the Minerva group, 
one in the 11882 (1990 RA3),
one in the Maria family (all groups near the Eunomia orbital region),
one was in the Astraea family, and the remaining asteroids were found
in families in the outer central main belt, including one in the 
Merxia family.   The fact that the majority
of V-type candidates belonging to a family are to be found in 
the orbital region of the Eunomia asteroid family may be an indirect
confirmation of the role that the parent body of this family may have
played as a source of V-type asteroids in the central main belt
(Carruba et al. 2009).

To numerically check for the statistical significance of the Eunomia and
Merxia/Agnia parent bodies as sources of basaltic material in the
central main belt, we perfomed this numerical experiment:
following the approach of Carruba and Machuca (2011), we 
estimated the probability that a number of 
objects be produced by a Poisson distribution assuming that the expected 
number of $k$ occurrence in a given interval is given by:

\begin{equation}
f(k,\lambda) = \frac{{\lambda}^k e^{-\lambda}}{k!},
\label{eq: poisson}
\end{equation}

\noindent
where $\lambda$ is a positive real number, equal to the expected number of
occurrences in the given interval.  For our purposes we used for $\lambda$ 
the mean values of objects expected in the regions of the halos of 
the Eunomia and Merxia/Agnia families\footnote{We did not perform
a statistical analysis of the region of Hansa, because this family 
is in a stable island located between the 3J:-1A and 8J:-3A mean-motion 
resonances in proper $a$, and between the ${\nu}_6$ and ${\nu}_5$ 
secular resonance in $\sin(i)$ (Carruba 2010b).  While  80\% of the 
highly inclined V-type photometric candidates are members of this family
this may not be statistically significant because of the local dynamics.}.
This is given by the equation:

\begin{equation}
\lambda = N_{ast}\cdot \frac{V_{halo}}{V_{Tot}}
\label{eq: weighted_mean}
\end{equation}

\noindent where $N_{ast}$ is the total number of V-type photometric
candidates, $V_{halo}$ is the volume in the $(a,e,sin(i))$ space
occupied by the family halo and $V_{Tot}$ is the total volume occupied
by all photometric candidates, defined as the region between 
the 3J:-1A and 5J:-2A mean motion resonances with Jupiter in 
semi-major axis (i.e., $2.508 < a < 2.826$~AU), and excentricity
and sin(i) from zero to the maximum value of any V-type candidate
in the central main belt (0.2865 and 0.3822, respectively). 
For the volume of the Eunomia family halo we used a simple parelelepipedal
defined according to the maximum values of $(a,e,sin(i))$ observed
for the Eunomia family core and halo as found in Carruba et al. (2007a): 
$2.508 < a < 2.707$~AU$,0.12 < e < 0.16, 0.18 < sin(i) <0.24$ (Eunomia 
core), and $2.508 < a < 2.753$~AU$,
0.12 < e < 0.19, 0.18 < sin(i) <0.26$ (Eunomia halo).  
Since we have no information
on the possible extent of the proto-family that originated
the Merxia and Agnia families, in this case we used as boundaries 
the combined limits of the halos of the two families in the
$(a,e,sin(i))$ space, as determined by Carruba et al. (2013b):
$2.707 < a < 2.826, e < 0.15$, and $\sin(i) <0.12$. 

Using standard statistical terminology, we define the null hypothesis 
as the possibility that the data are drawn from a
given distribution.  We can reject the null hypothesis if it is 
associated with a probability lower than a threshold, 
usually of the order of 1\%.  Using Eq.~\ref{eq: poisson}, 
we found that the probability that the 6 asteroids
in the Eunomia core region, 15 asteroids in the Eunomia halo
region, and the 15 found in the Merxia/Agnia halo region could be explained
as fluctuations of a Poissonian distribution are of $7.0 \cdot 10^{-3}, 
1.7\cdot 10^{-4}$ and $8.1 \cdot 10^{-3}$, i.e., 
below the null hypothesis threshold ($1.0\cdot 10^{-2}$).  A similar
analysis done for the uniform and Gaussian distributions, according
to the procedure described in Carruba and Machuca (2011), also provided
values of probabilities below the null hypothesis, so suggesting
that the higher density of objects found in the Eunomia and Merxia/Agnia
regions may be caused by a possible local source.

Finally, since these 1-dimensional statistical analysis are 
somewhat dependent on the choice of the regions boundaries, we also perfomed 
Mardia's test (Mardia 1970) on multivariate normality for the whole 128 
asteroid population.  We assume that the distribution in the 
$(a,e,\sin(i))$ domain has a tri-variate Gaussian probability density 
function given by:

\begin{equation}
f({\bf x}) = \frac{1}{(2 \pi)^{3/2} \sqrt{\left| {\Sigma} \right|}}
  \exp{(-\frac{1}{2}({\bf x}-\mu)'{\Sigma}^{-1}({\bf x}-\mu))}, 
\label{eq: prob_tri}
\end{equation}
  
\noindent
where ${\bf x}$ is the vector $(a,e,\sin{(i)})$, $\mu$ is the vector with
components equal to the mean values of $a$, $e$ and $\sin{i}$ of the observed
population, and $\Sigma$ is the covariance matrix, 
that can be estimated numerically having $n$ {\bf $x_j$} vectors using the 
equation:

\begin{equation}
\Sigma = \frac{1}{n} \sum_{j=1}^{n} ({\bf x_j}-\mu)\cdot({\bf x_j}-\mu)^{T},
\label{eq: covariance}
\end{equation}
Mardia's test is based on multivariate extensions of skewness and kurtosis 
measures. For a sample $(x_1, ..., x_n)$ of p-dimensional vectors we can compute
the A and B parameters given by:

\begin{equation}
A=\frac{1}{6n} \sum_{i=1}^{n} \sum_{j=1}^{n} \left[({\bf{x}}_i-\mu)^{T}
    {\Sigma}^{-1} ({\bf{x}}_j-\mu)\right]^3,
\label{eq: A_mardia}
\end{equation}

\noindent and

\begin{equation}
B = \frac{\sqrt{n}}{\sqrt{8p(p+2)}}\left[\frac{1}{n}\sum_{i=1}^{n}
  \left[({\bf{x}}_i-\mu)^{T}{\Sigma}^{-1} ({\bf{x}}_i-\mu) \right]^2-p(p+2)\right],
\label{eq: B_mardia}
\end{equation}

\noindent where ${\Sigma}^{-1}$ is the inverse of the 
covariance matrix given by Eq.~\ref{eq: covariance}. Under the null 
hypothesis of multivariate normality, the statistic $A$ will have 
approximately a chi-squared distribution with $\frac{1}{6}p(p + 1)(p + 2)$ 
degrees of freedom (10 for p =3), and $B$ will be approximately 
standard normal with mean zero and standard deviation one.  We computed $A$
and $B$ for our sample of 128 asteroids and we obtained $A = 27.389$ and 
$B = 1.7 \cdot 10^{3}$, that have probabilities of 
being associated with a $\chi^{2}$ 
and a normal distribution lower than $4.5 \cdot 10^{-5}$.  
We can therefore safely
assume that V-type photometric candidates do not follow a tri-variate
Gaussian distribution as a whole.

In the next section we will further analyze the six dynamical regions
defined in this section.

\section{Groups of V-type candidates and their possible origin}
\label{sec: V-type_groups}

First, to test for the reality of the six concentration groups identified 
in Sect.~\ref{sec: taxonomy-sdss} we performed the following numerical
experiment: we computed the distances with respect to (15) Eunomia according
to the distance metrics (Zappal\'{a} et al. 1995)~\footnote{We should point 
out that these distances, while expressed in $m/s$, are not exactly 
ejection velocities, that should be computed using Gauss equation (see
also Carruba et al. 2003).
Since, however, ejection velocities computed with such an approach
also depend on the unknown true anomaly and argument of pericenter
at the moment of impact of the parent body, in this work we prefer
to use this simpler approach.}:

\begin{equation}
d=na \sqrt{h_1\left({\frac{\Delta a}{a}}\right)^2 + 
h_2{(\Delta e)}^2 +h_3{(\Delta (sin(i)))}^2},
\label{eq: metric_g-s}
\end{equation}

\noindent where $n$ is the asteroid proper motion, $a, e,$ and $sin(i)$ are
the standard set of proper elements (semi-major axis, eccentricity, 
and sine of the inclination), and $h_i$ (i=1,2,3) are numerical constants
equal to 5/4, 2, and 2 for the standard metric of Zappal\'{a} et al. (1995).
Other choices of the $h_i$ constants are possible and discussed in the 
literature (see also Carruba and Michtchenko 2007).  The advantage of
using distances $d$ rather than bidimensional plots as those shown 
in Fig.~\ref{fig: V-type_central} is that these quantities account 
for distances in a three-dimensional space and are not
subjected to the distortions that arise from plotting three dimensional
distributions in two-dimensional spaces.

\begin{figure*}

  \centering
  \begin{minipage}[c]{0.5\textwidth}
    \centering \includegraphics[width=3.5in]{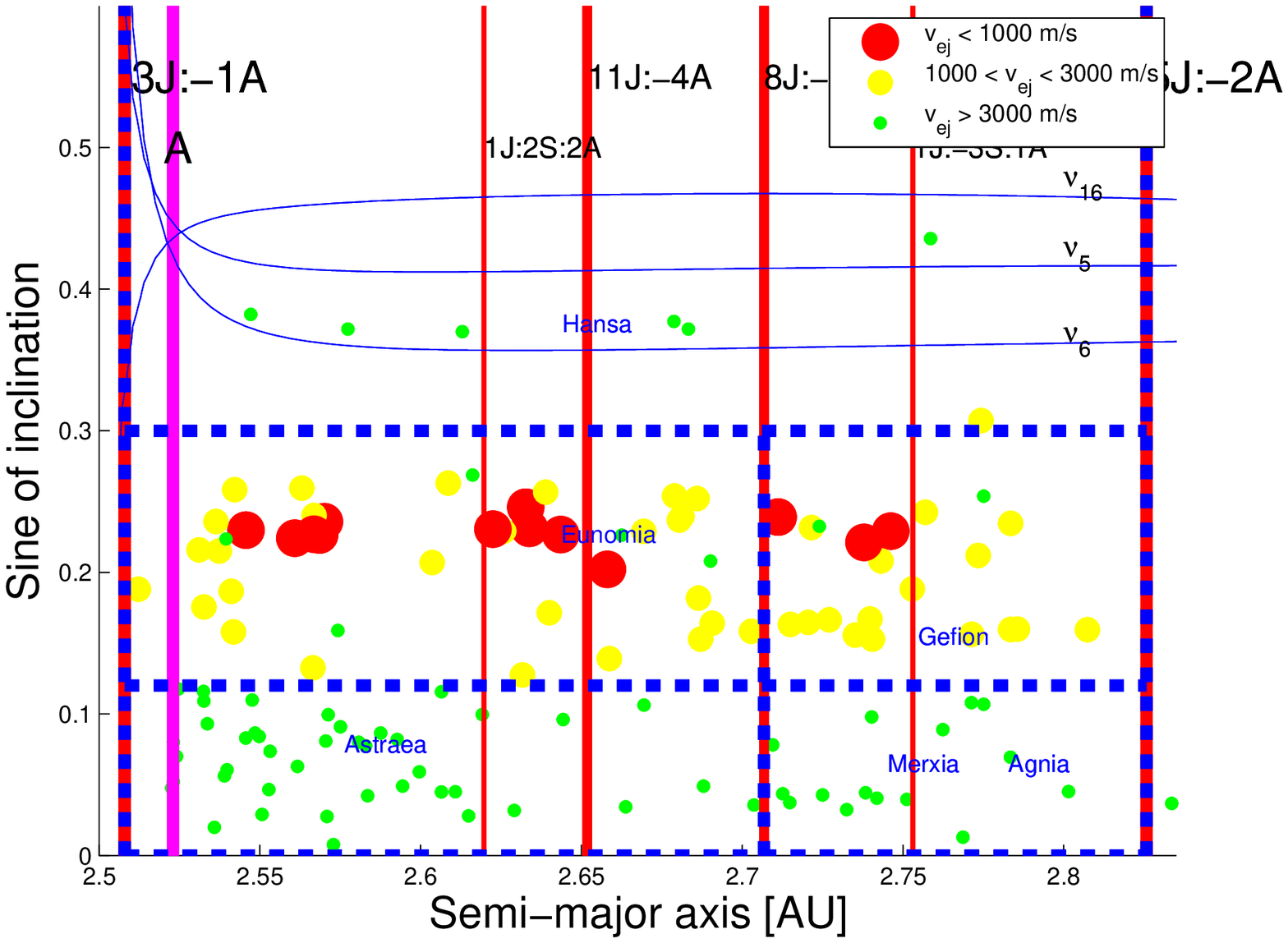}
  \end{minipage}%
  \begin{minipage}[c]{0.5\textwidth}
    \centering \includegraphics[width=3.5in]{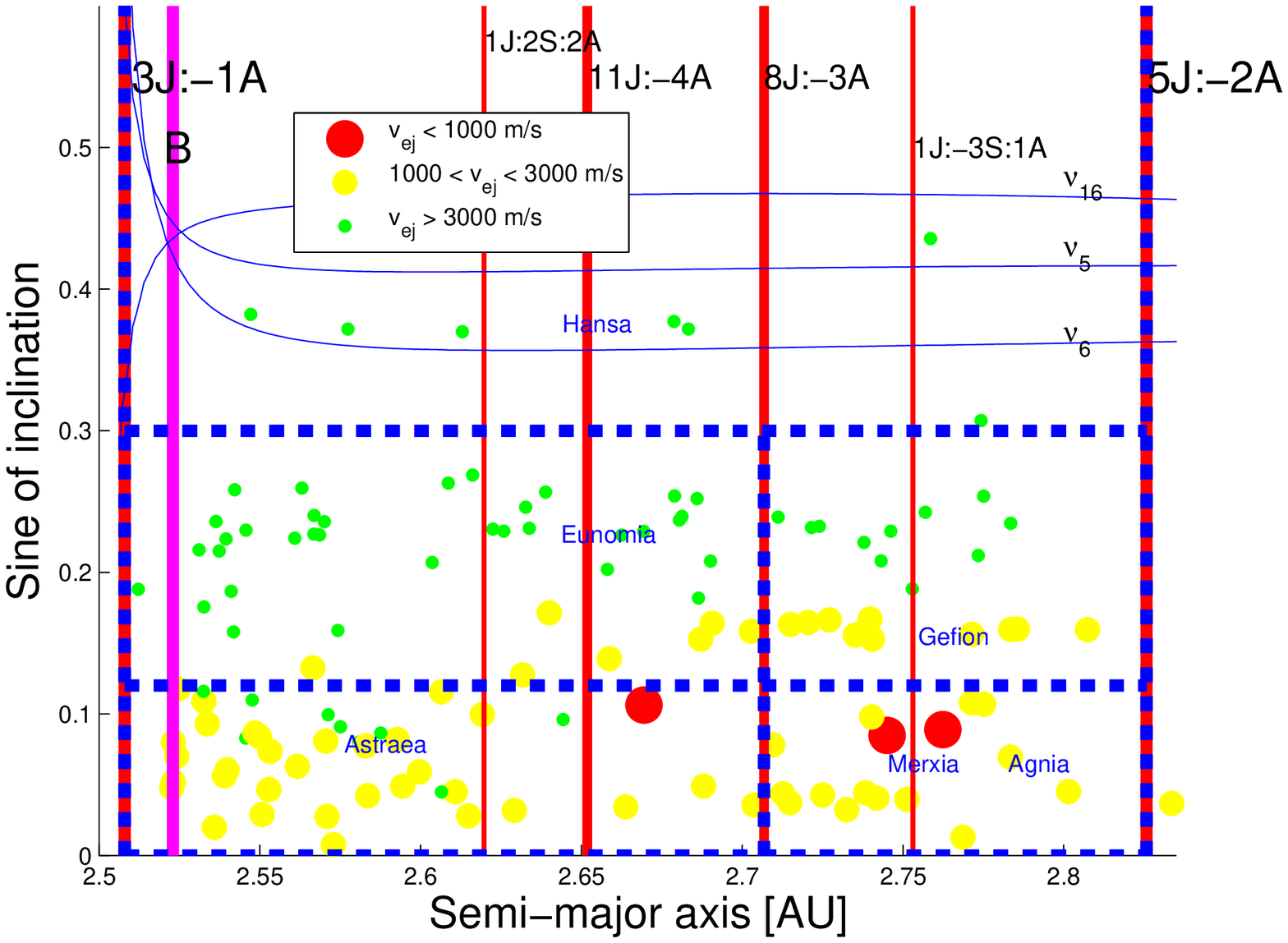}
  \end{minipage}

\caption{An $(a,sin(i))$ plot of V-type photometric candidates 
in the central main belt.  Distances in proper element domains 
of V-type asteroids
in the central main belt with respect to (15) Eunomia are
plotted with this color code:  red full dots
identify objects with $d < 1000$~m/s, yellow full dots asteroids
with $1000 < d < 3000$ m/s, and green dots bodies with $d > 3000 m/s$.
The other symbols have the same meaning as those in 
Fig.~\ref{fig:  V-type_central}, panel B.}
\label{fig: ej_vel}
\end{figure*}

Fig.~\ref{fig: ej_vel}, panel A, displays an $(a,sin(i))$ plot of V-type 
photometric candidates in the central main belt.  Distances in 
proper element domains of V-type asteroids
in the central main belt with respect to (15) Eunomia are
plotted with this color code:  red full dots
identify objects with $d < 1000$~m/s, yellow full dots asteroids
with $1000 < d < 3000$ m/s, and green dots bodies with $d > 3000 m/s$.
The other symbols are the same as in Fig.~\ref{fig:  V-type_central}.
Objects nearest to the Eunomia family (red and yellow dots) correspond
to the Eunomia family region and the two horizontal ``strips'' in the
outer central main belt at $sin(i) > 0.12$ defined in 
Sect.~\ref{sec: taxonomy-sdss}.  Green asteroids are found
in the ``Astraea'' region, in the $sin(i) < 0.12$ outer central main
belt strip, and in the highly inclined region.  With the exception of
seven objects, we confirm the $0.12 < sin(i) < 0.3$ criteria for asteroids
possibly originating from the Eunomia family defined in  
Sect.~\ref{sec: taxonomy-sdss}. 

We also computed distances with respect to the Merxia ``region''.  Since
the original orbital parameters of the possible parent body of the 
Merxia and Agnia families are unknown, we computed distances with 
respect to the largest surviving fragment, (808) Merxia itself.
Our results are shown in Fig.~\ref{fig: ej_vel}, panel B.  Low-inclined 
objects have smaller distances with respect to Merxia than Eunomia.
The strip of aligned asteroids in the Gefion region, that was found to be
at relatively low distances from (15) Eunomia, appears to be within reach
of (808) Merxia as well, which suggests that either sources could be 
responsible for these asteroids.  Overall, these results 
seem to suggest a second possible source of basaltic material in the 
central main belt.

We then turned our attention to the diameters of photometric V-type 
candidates in the central main belt. Results from the Wide-field Infrared 
Survey Explorer (WISE) (Wright et al. 2010), and the NEOWISE 
(Mainzer et al. 2011) enhancement to the WISE mission recently 
allowed to obtain diameters and geometric albedo values for more 
than 100,000 Main Belt asteroids (Masiero et al. 2011).  Of the 
127 V-type candidates in the central main belt, 27 had values
of diameters and geometric albedos listed in the WISE dataset.
Concerning the other asteroids, their diameters and albedos
can be estimated using the relationship (Harris and
Lagerros 2002):

\begin{equation}
D= \frac {1329 km}{\sqrt{p_V}} \cdot 10^{-H/5},
\label{eq: D_H_pv}
\end{equation}

For the objects lacking WISE albedo data, we used the mean value of 
the albedo available for the other 27 bodies, i.e., $p_V = 0.238$.
The minimum value of albedo was of 0.0518, and the maximum of 
0.4268.  Each of the extreme values of $p_V$ was only reached once, 
and if we eliminate such outliers from the mean, we obtained a mean 
value of 0.2379, very close to the mean value obtained with the 
outliers.

\begin{figure}

  \centering
  \centering \includegraphics [width=0.45\textwidth]{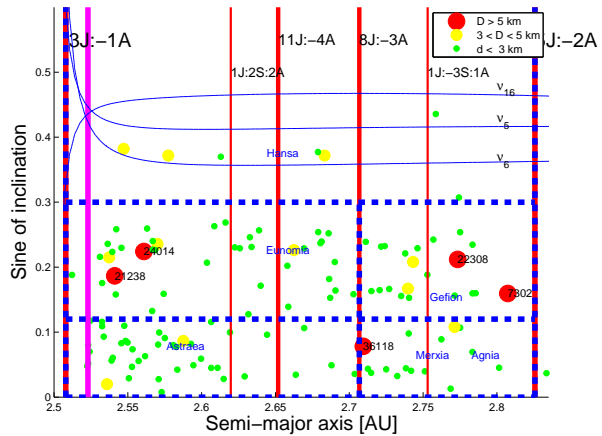}

\caption{An $(a,sin(i))$ projection of V-type photometric candidates
in the central main belt.  The size of the symbol is associated
with the asteroid diameter, according to the figure legend.  
The other symbols have the same meaning as those in 
Fig.~\ref{fig:  V-type_central}, panel B.} 
\label{fig: V-sizes}
\end{figure}

Fig.~\ref{fig: V-sizes} displays an $(a,sin(i))$ projection 
of V-type photometric candidates in the central main belt.  The size 
of the symbol is associated with the asteroid estimated diameter:
large full red dot display the orbital position of the five objects
with $D > 5$ km, yellow full dots show the position of the asteroids
with $3 < D < 5$~km, and the green dots are associated with 
smaller bodies.  One can notice that, with the exception of
the asteroid (36118), that is barely 5 km in diameter, all
larger objects are found at $sin(i) > 0.12$. Only two 
medium-sized asteroids are encountered in the ``Astraea'' region.

To estimate how much mass is contained in the currently known V-type
candidates in the central main belt we estimated the mass
of each object using the equation (Moskovitz et al. 2008a, eq. 6):

\begin{equation}
M(H) = (1.28 \cdot 10^{18} kg) \frac{\rho}{{p_V}^{3/2}} 10^{-0.6H}, 
\label{eq: mass_mosko}
\end{equation}

\noindent  where $\rho$ is the bulk density, assumed
equal to 3000 $kg/m^3$ for V-type objects, and asteroids have been assumed
to be spherical bodies.  Using the values of geometrical albedo
$p_V$ previously discussed in this section, we obtain a total mass
of $1.81\cdot 10^{15}$~kg , which is just 0.139\% of the estimated 
mass escavated from craters in Vesta $1.3 \cdot 10^{18}$~kg (Moskovitz 
et al. 2008a).  Only a very minor fraction of the basaltic material
present in the main belt is therefore located in the central main belt.
In the next section we will discuss how the V-type photometric
candidates are located with respect to the local web of mean-motion
and secular resonances.

\section{Dynamical Maps}
\label{sec: dyn_maps}

To investigate in further detail the local dynamics we computed
a dynamical map of synthetic proper elements with 19800 test particles
in the region of the central main belt.
We used the SWIFT\_MVFS symplectic 
integrator from the SWIFT package (Levison and
Duncan, 1994) modified by Bro\v{z} (1999) so as to include on-line
digital filtering to remove all frequencies with periods less than 600
yr.  We used a step in $a$ of 0.005 AU and in $i$ of $0.2^{\circ}$, and 
took particles in an equally spaced grid of 165 by 120
 particles in the $(a,sin(i))$ plane, the representative plane
for studying diffusion of members of asteroid families
\footnote{Our particles covered
a range between 2.5 and 2.828 AU in $a$, and $0^{\circ}$ and $23.8^{\circ}$ 
in $i$, respectively.  Since the local dynamics has been studied in detail
in several other works, (see also Carruba et al. 2007b), and since 
the proper excentricity of asteroid is changed more easily then its proper 
inclination, we did not performed in this work analysis in the 
$(a,e)$ and $(e,sin(i))$ planes.}.  The initial 
values of $e, \Omega, \omega,$ and $\lambda$ of the test particles
were fixed at those of (15) Eunomia, the largest member of its family
and a possible source of V-type asteroids, according to Carruba et al. 
(2007b).  We computed synthetic proper elements $(a, e, sin(i))$ 
and frequencies $(n,g,s)$ of these test particles with the approach 
described in Carruba (2010).

\begin{figure*}

  \centering
  \centering \includegraphics [width=0.95\textwidth]{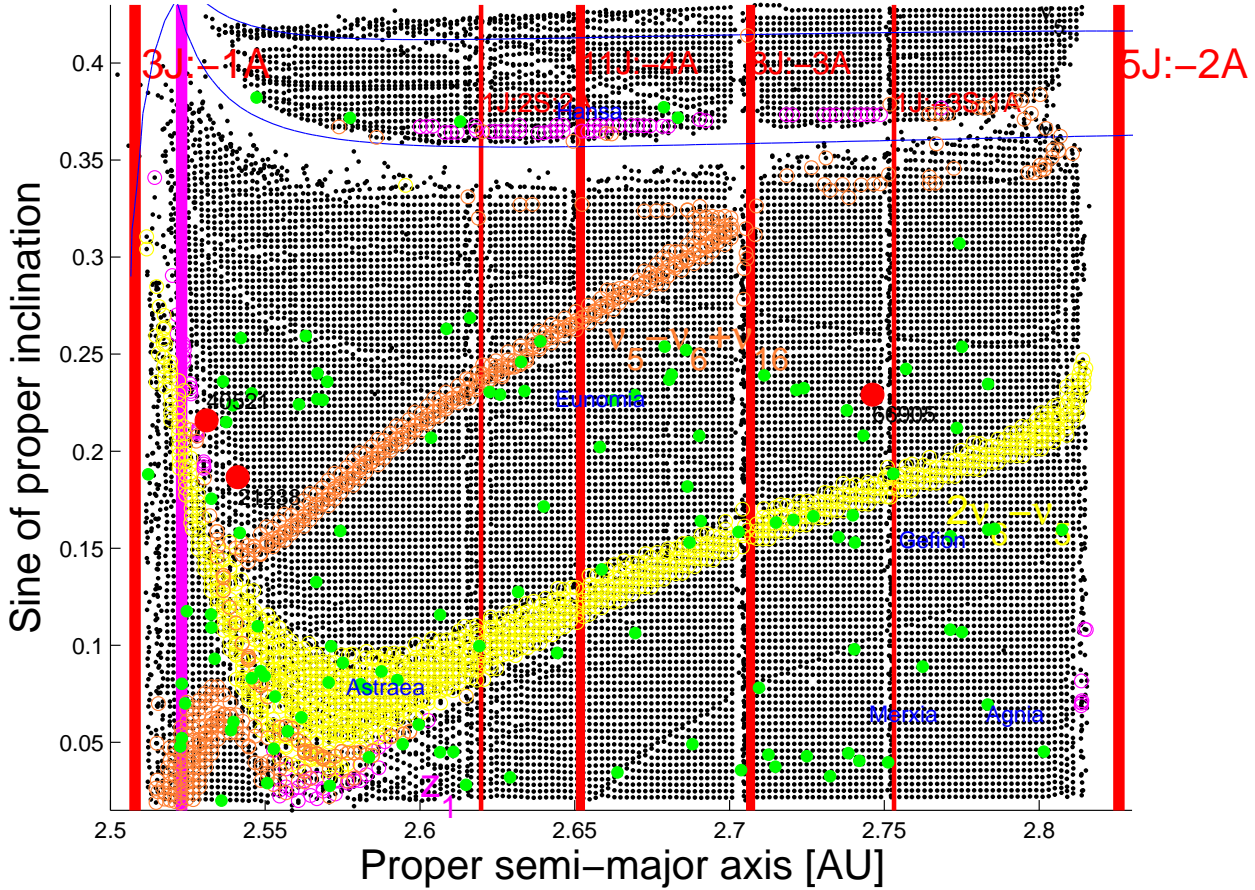}

\caption{An $(a,sin(i))$ proper element map of the central main belt.
Black dots identify
the orbital position in the plane of proper $(a, sin(i))$ of each simulated
test particle. Color circles are associated with asteroids likely 
to be in secular resonance configurations, according to the color
code described in the text. The other symbols have the same meaning as in 
Fig.~\ref{fig:  V-type_central}.} 
\label{fig: map_ai-cen}
\end{figure*}

Results are shown in Fig.~\ref{fig: map_ai-cen}.  Black dots identify
the orbital position in the plane of proper $(a, sin(i))$ of each simulated
test particle.  Unstable regions will appear as devoided of test particles,
mean-motion resonances will show as vertical alignements, and secular
resonances will be associated with inclined alignements of dots.
In color we show test particles whose frequency values where within
$\pm 0.3~arcsec/yr$~\footnote{The so-called 
likely resonators, or objects with a probability higher than 90\%
of being in librating states of the resonances, as defined in Carruba 
(2009a).  Likely resonators are found by equating the values of the 
asteroidal frequencies to the planetary ones.  In the case
of the $z_1$ secular resonance, we have $g-g_6+s-s_6 \simeq 0$, 
which implies $g+s = g_6+s_6 =1.898 arcsec/yr$.} from the center 
of the $z_1 = {\nu}_6+{\nu}_{16} = g-g_6+s-s_6$ (magenta dots),
the $2{\nu}_6-{\nu}_5 = g+g_5-2g_6$ (yellow dots), and from the 
${\nu}_5 -{\nu}_6+{\nu}_{16}= g+g_5-2g_6$ (orange dots) resonances, 
respectively.
The other symbols are the same as in Fig.~\ref{fig:  V-type_central}.
While all the other secular resonances up to order six present in the 
region were studied, we choose to display in the figure the orbital 
location of these three resonances only because of their dynamical importance.
The $z_1$ secular resonance played a major role in the dynamical evolution
of the Agnia (Vokhrouhlick\'{y} et al. 2006) and Padua (Carruba 2009a) 
families, that are characterized by the fact that the majority of 
their members are in librating states of this resonances. The $z_1$ conserved 
quantity allowed to obtain better estimates of the 
extent of the original ejection velocity field of these two families.
The $2{\nu}_6-{\nu}_5$ had one of the largest population of 
likely resonators in our dynamical map, and its possible role
in the dynamical evolution of V-type photometric candidates
will be further investigated in this section.  Finally, the 
${\nu}_5 -{\nu}_6+{\nu}_{16}$ secular resonance was shown to be one 
possible mechanism of delivery of basaltic material from the Eunomia family
to lower inclination regions in Carruba et al. (2007a).  Its dynamical 
importance will also be discussed in Sect.~\ref{sec: yarko}.

Other minor secular resonances were identified in the dynamical map, but 
were not shown for simplicity.   The Eunomia and Gefion regions
are crossed by the $3{\nu}_6-{\nu}_5$ and $z_2 = 
2{\nu}_6+{\nu}_{16} = 2(g-g_6)+s-s_6$ non-linear secular resonances.
The $2{\nu}_6+{\nu}_{17} = 2(g-g_6)+s-s_7$ secular resonance is present
in the Merxia region.  Finally, the Hansa region is crossed
by the ${\nu}_5+2{\nu}_{16} = g-g_5+2(s-s_6), {\nu}_6-{\nu}_{16}= g-g_6
-s+s_6$, and ${\nu}_5-{\nu}_{16}= g-g_5-s+s_6$, secular resonances 
(see also Carruba 2010).  The last entry of Table~\ref{table: V_carvano}
identifies asteroids that are likely resonators of the discussed secular
resonances.  Incidentally, there are three
objects (93620, 108901, 225791) that are likely resonators of two resonances
at the same time, i.e., they are at the crossing of two secular resonances.

\begin{figure}

  \centering
  \centering \includegraphics [width=0.45\textwidth]{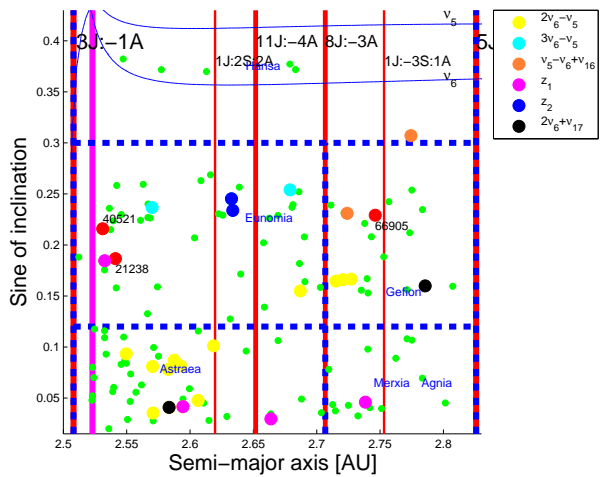}

\caption{Projection in the proper $(a,sin(i))$ plane 
of the likely resonators population (colored circles) 
listed in Table~\ref{table: res_V}, according to the color code
displayed in the figure legend.} 
\label{fig: V_sec_res}
\end{figure}

To understand the importance of secular dynamics for the evolution
of V-type asteroids in the central main belt, we also selected 
``likely resonators'' among the photometric V-type candidates. 
Our results are listed in Table~\ref{table: res_V}, that displays
the name of the resonance, its type, i.e., what asteroidal frequencies
are involved, the resonance center as previously defined in this
section, and the number of likely resonators for the resonances
that had at least one librating candidate.  An $(a,sin(i))$ projection
of the identified likely resonators is also shown in 
Fig.~\ref{fig: V_sec_res}.

%%%%%%%%%%%%%%%%%%%%%%%%%%%%%%%%%%%%%%%%%%%%%%%%%%%%%%%%%%%%
\begin{table*}
\begin{center}
\caption{{\bf Likely resonators among V-type photometric candidates.}}
\label{table: res_V}
\vspace{0.02cm}
\begin{tabular}{r l c c c }        % centered columns (8 columns)
\hline\toprule 
Resonant name & Type  & Resonant center [arcsec/yr]& \# of likely resonators \\
\midrule %\hline
%%%%%%%%%%%%%%%%%%%%%%%%%%%%%%%%%%%%%%%%%%%%%%%%%%%%%%%555
$2{\nu}_6-{\nu}_5$          & $g$    & 52.229 & 15 \\
$3{\nu}_6-{\nu}_5$          & $g$    & 40.236 &  3 \\
${\nu}_5-{\nu}_6+{\nu}_{16}$ & $s$    &-50.300 &  2 \\
${\nu}_6+{\nu}_{16} = z_1$   & $g+s$  &  1.998 &  4 \\
$2{\nu}_6+{\nu}_{16} = z_2$  & $2g+s$ & 30.141 &  2 \\
$2{\nu}_6+{\nu}_{17}$        & $2g+s$ & 53.141 &  2 \\
\bottomrule%\hline
\end{tabular}
\end{center}
\end{table*}

The largest population of likely resonators is found in the 
$2{\nu}_6-{\nu}_5$ secular resonance, with 15 candidates.  All the other
resonances had four resonators or less.  The relatively large number
of candidates in this $g$-type resonance may have interesting
repercussions on the dynamics of V-type asteroids.  One may wonder
if the cluster of V-type asteroids in the Astraea region may be 
caused by an accumulation of asteroids in this region.
To start answering this question, following the approach of 
Carruba et al. (2013b) we integrated the likely resonators
in the $2{\nu}_6-{\nu}_5$ secular resonances for 10 Myr under
the influence of all planets, and checked the behavior of
the resonant argument.  11 asteroids are in librating states,
2 in circulating states, and 2 are alternating phases of circulation
and libration.  The large fraction of actual resonators
among the candidates suggest that this is indeed a powerful resonance.
Its role when the Yarkovsky force is considered will be discussed
in Sect.~\ref{sec: yarko}, but, being a pericenter resonance,
it seems unlikely that it will be able to significantly change the 
inclination of wandering asteroids.  In the next section we 
will try to answer this and other questions on the possible origin 
of photometric V-type asteroids in the central main belt.

\section{Yarkovsky evolution}
\label{sec: yarko}

To investigate the dynamical evolution of V-type candidates in the
central main belt, we integrated clones of these objects
with SWIFT-RMVSY, the symplectic integrator of Bro\v{z} (1999) that 
simulates the diurnal and seasonal versions of the Yarkovsky effect, 
over 30 Myr and the gravitational influence of all planets from Venus 
to Neptune (Mercury was accounted for as a barycentric correction in 
the initial conditions).  We used values of the Yarkovsky parameters 
appropriate for V-type asteroids: a thermal conductivity 
$K = 0.001 W/m/K$, a thermal capacity $C = 680~J/kg/K$, surface density 
1500 $kg/m^3$, a Bond albedo of 0.1, a thermal emissivity of 0.95, and a 
bulk density of 3000 $kg/m^3$.  As our goal is to investigate the maximum
possible diffusion of asteroids, we used two sets of spin axis orientations 
with $\pm90^{\circ}$ with respect to the orbital plane, since these 
obliquities maximize the speed of the Yarkovsky effect.  We assumed 
periods obtained under the approximation that the rotation frequency is
inversely proportional to the object's radius, and that a 1 km
asteroid had a rotation period of 5 hours (Farinella et al. 
1998)~\footnote{Other choices of rotation periods are possible.
One can choose a distribution of rotation frequencies similar to that
of other families, and randomly choose values for each asteroid.
However, Cotto-Figueroa et al. (2013) have shown that the 
YORP effect is extremely sensitive to the topography of the asteroid, 
and its small changes.  We therefore believe that 
in the end it may make little difference what initial rotation period 
is chosen.  The rotation period of an asteroid will change during a
YORP cycle in ways that are not currently well understood.
Since our goal in this section is to preliminary
investigate the fraction of surviving resonators when non-gravitational
forces are considered, we believe that our simpler approach is  
justified.}.  No re-orientations were considered, so that the drift 
caused by the Yarkovsky effect was the maximum possible.
 
We first performed ``fast simulations'' of these asteroids, assuming
that all objects had $100~m$ diameters, over 30 Myr.  
Objects so small will drift 
much faster than the real asteroids, allowing for quicker simulations 
time-scales.  On the other end, details of evolution in higher order
mean-motion and secular resonances may be lost because of the faster
drift.  To account for this problem, we will also perform simulations
with the real larger values of diameters ``slow simulations'', 
scaled by factors ${\cos{30^{\circ}}}$ and ${\cos{60^{\circ}}}$ to account 
for values of the spin obliquity other than zero, that we will discuss 
later on in this section.

\subsection{Results of the ``fast simulations''}
\label{sec: fast_sims}

\begin{figure}

  \centering
  \centering \includegraphics [width=0.45\textwidth]{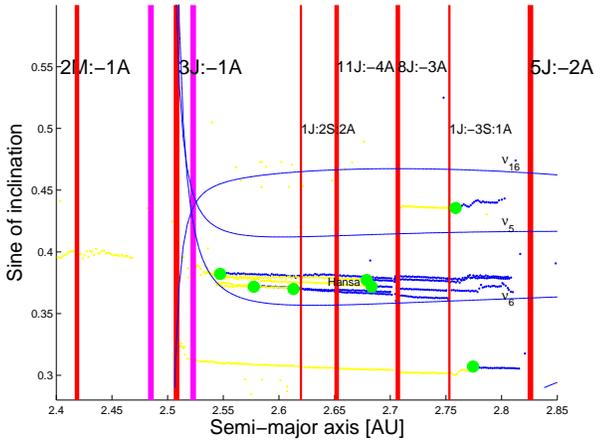}

\caption{An $(a,sin(i))$ projection of the time evolution of fast 
clones or real V-type candidates in the Hansa region. Blue dots represent
snaphsots of the orbital evolution (sampled at each 1.2 Myr) of clones
of real asteroids with $0^{\circ}$ obliquity, while yellow dots are associated
with clones with $180^{\circ}$ obliquity.  The other symbols are the same 
as in Fig.~\ref{fig: ej_vel}.}
\label{fig: hansa_region}
\end{figure}

We obtained synthetic proper elements every 1.2 Myr 
for all the simulated asteroids with
the approach described in Kne\v{z}evi\'{c} and Milani (2003), modified
as in Carruba (2010).  Fig.~\ref{fig: hansa_region} displays our
results for asteroids in the Hansa region.  Blue dots represent
snaphsots of the orbital evolution of clones
of real asteroids with $0^{\circ}$ obliquity, while yellow dots are associated
with clones with $180^{\circ}$ obliquity. 
The magenta lines display the chaotic layer near the 3J:-1A 
studied by Guillens et al. (2002), as defined in Morbidelli and 
Vokrouhlick\'{y} (2003).  The other symbols are the same 
as in Fig.~\ref{fig: ej_vel}.

One object near the Watsonia family was included in this region, having
$\sin(i) > 0.3$, the criteria used by Gil-Hutton (2006) for 
identifying objects of high-$i$.  However, since its inclination is lower 
than the central value of the ${\nu}_6$ secular resonance, other authors
(Carruba 2010) may not consider this object a highly inclined body.
The clone of the asteroid in the Gallia family region reached Mars
crossing values of eccentricity when passing the 8J:-3A and was 
lost before the end of the simulation.
One may notice that asteroids in the Hansa region are able to cross
the 11J:-4A and 8J:-3A mean-motion resonances with only minor changes of
inclination and reach the region of the Tina and Gallia families.  Of particular
interest was a clone with $180^{\circ}$ obliquity, that succeded in crossing
the 3J:-1A mean-motion resonance.  This suggests that evolution in 
the other direction may also be possible as suggested by Roig et al. (2008), 
and may have been responsible
for the origin of the asteroids in the Hansa region, that may 
therefore be actual
Vestoids, and originated from the inner main belt.  A possible origin from 
the low-inclination central main belt seems unlikely, given the difficulties
in crossing the dynamical boundary caused by the presence of the ${\nu}_6$ 
secular resonance.
Additional studies are needed on this subject to prove this
hypothesis.  Obtaining spectra of these asteroids and performing a comparative
mineralogical analysis with respect to known Vestoids may provide important
clues in understanding the origin of these objects.

\begin{figure*}

  \centering
  \begin{minipage}[c]{0.5\textwidth}
    \centering \includegraphics[width=3.5in]{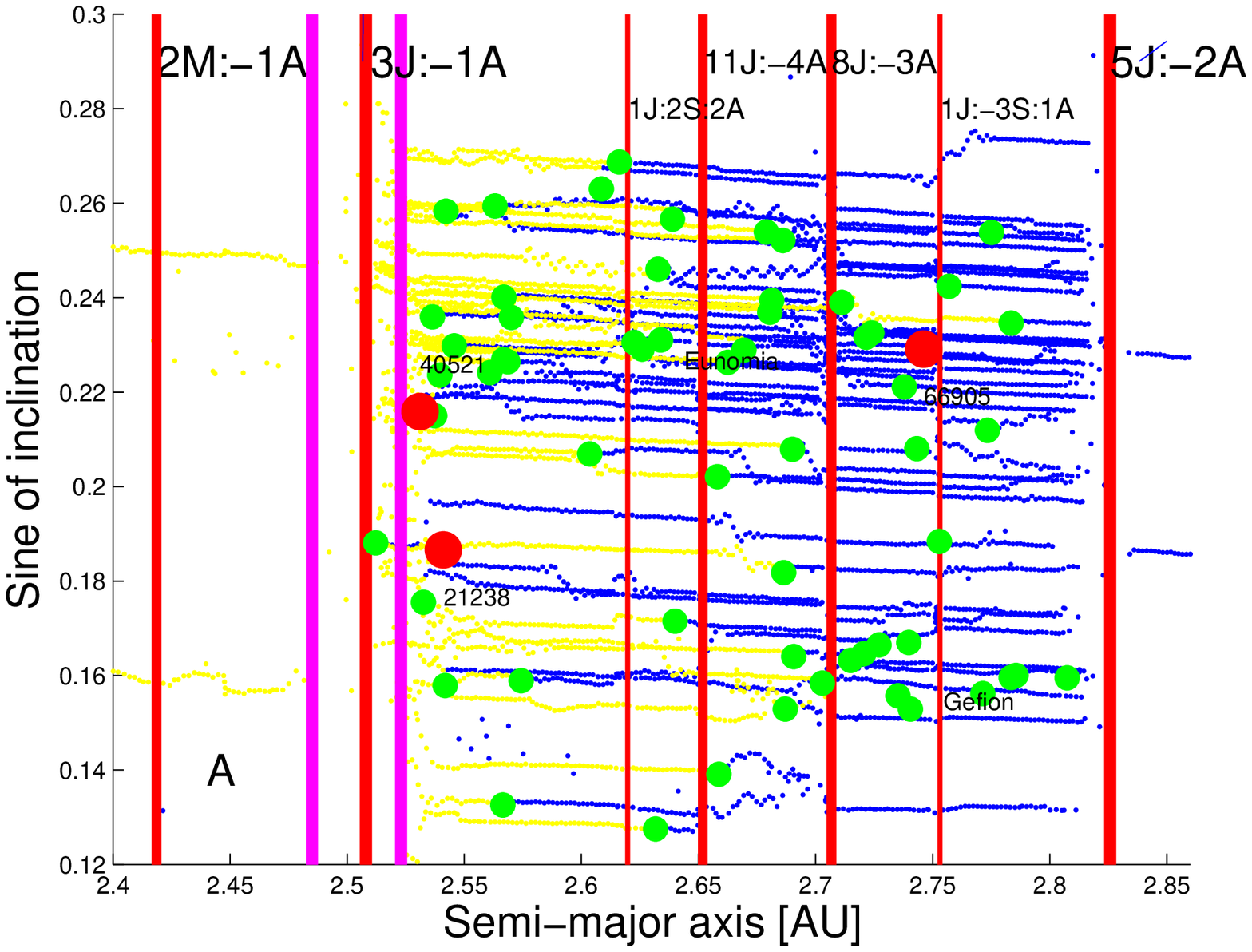}
  \end{minipage}%
  \begin{minipage}[c]{0.5\textwidth}
    \centering \includegraphics[width=3.5in]{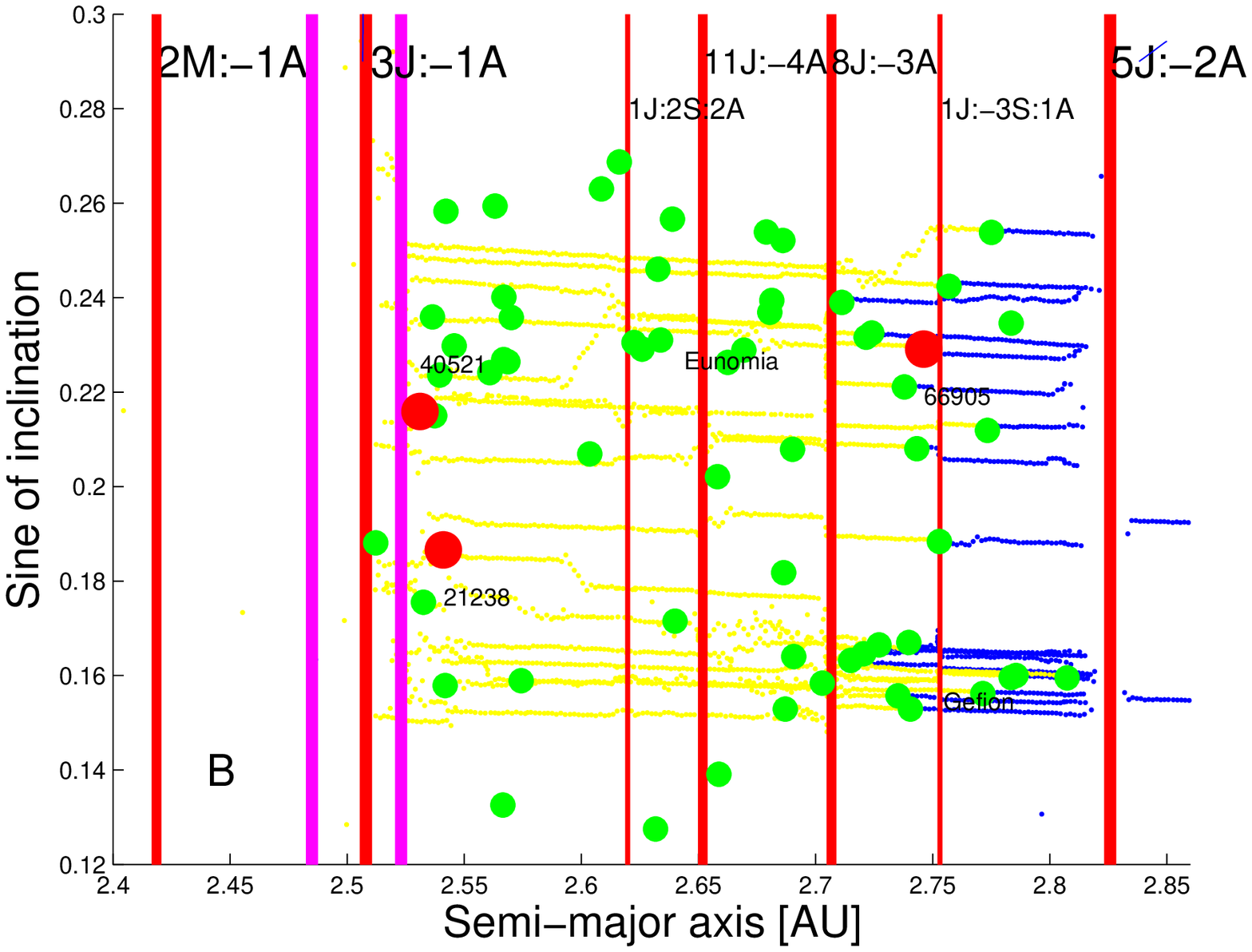}
  \end{minipage}

\caption{Panel A: An $(a,sin(i))$ projection of the time evolution of fast 
clones or real V-type candidates in the Eunomia region.  Panel B: the same for
clones of asteroids in the Gefion region.  See caption of 
Fig.~\ref{fig: hansa_region} for a description of the used symbols.}
\label{fig:  eun_gef_regions}
\end{figure*}

In Fig.~\ref{fig:  eun_gef_regions} we show the dynamical paths of clones
of particles in the Eunomia region (panel A), and in the Gefion area (panel B).
All three confirmed V-type asteroids in the central main belt are found in this
region.  Once again, we found particles (four) 
that manage to cross the main dynamical
barriers of the 3J:-1A and 5J:-2A mean motion resonance, suggesting that 
small objects comunication between inner and central main belt might be
possible (how effective is this mechanism for larger asteroids will be
discussed in Sect.~\ref{sec: slow_sims}).  This could also have 
implication on the origin of V-type asteroids in the inner outer
main belt (the area between the 5J:-2A and 9J:-4A mean-motion resonances 
with Jupiter), such as (10537) (1991 RY16), that could 
possibly be fragments of the parent body of the Merxia/Agnia families.
Also, we found 
that most particles can easily cross the 11J:-4A and 8J:-3A mean-motion 
resonances, and that migration from the Eunomia region to the Gefion one
(and vice-versa) is possible and could explain the presence of
objects such (66905), as discussed in Carruba et al. (2007a).   A few particles
were captured in the secular resonances described in 
Sect.~\ref{sec: dyn_maps} and experienced moderate changes in inclinations.
We did not however observed any particle able to reach the Astraea and 
Merxia regions in this simulation.

\begin{figure*}

  \centering
  \begin{minipage}[c]{0.5\textwidth}
    \centering \includegraphics[width=3.5in]{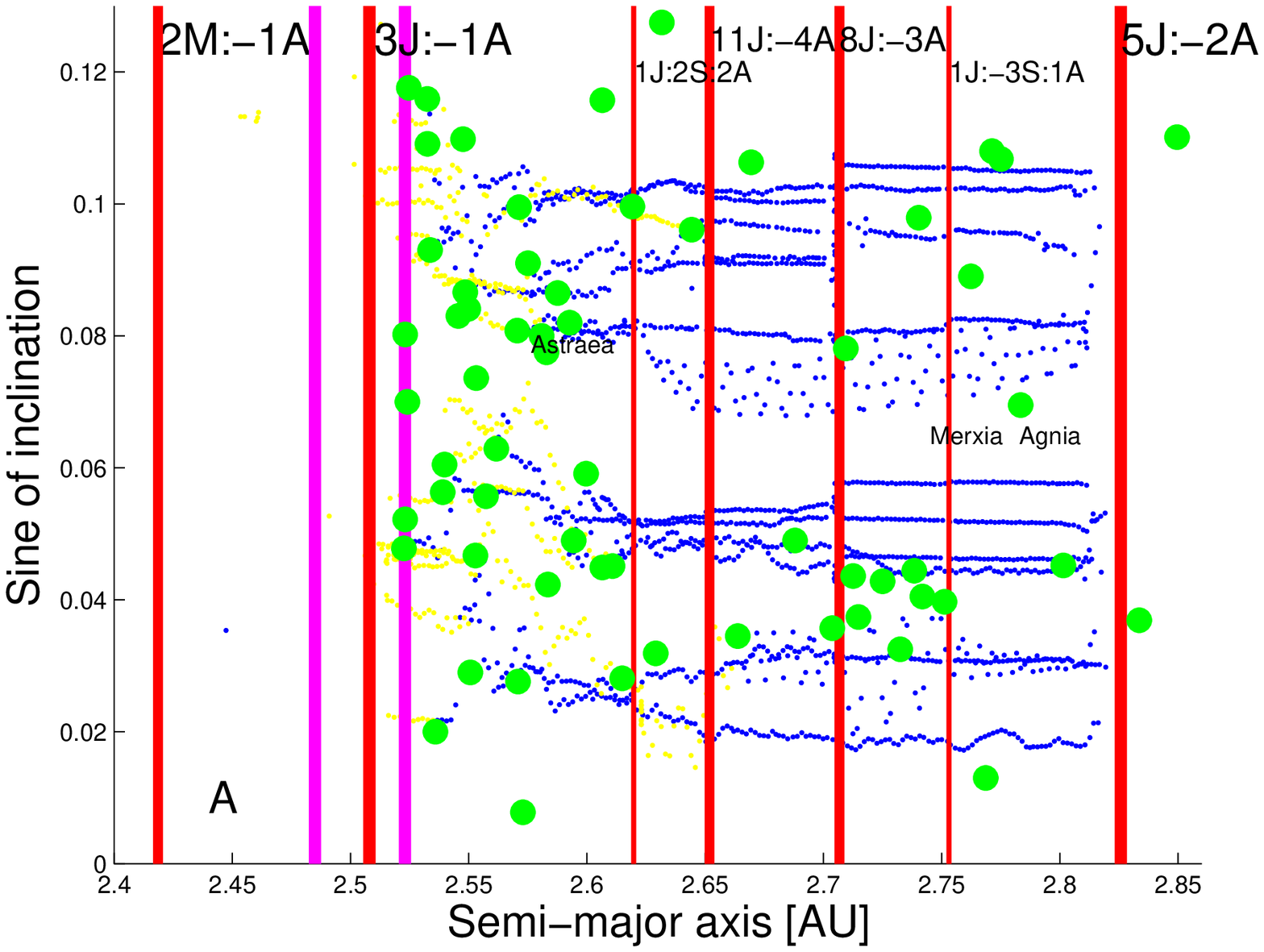}
  \end{minipage}%
  \begin{minipage}[c]{0.5\textwidth}
    \centering \includegraphics[width=3.5in]{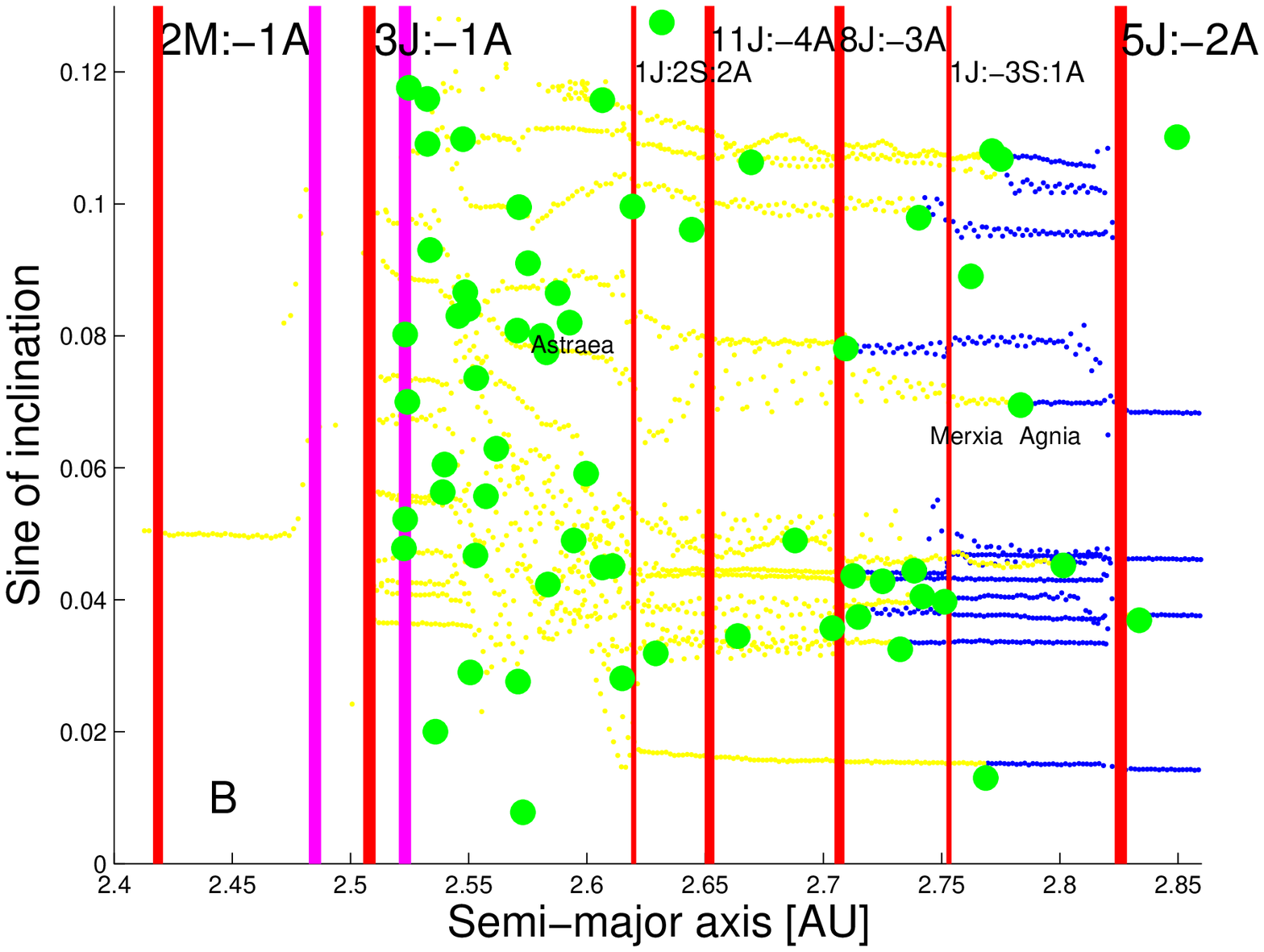}
  \end{minipage}

\caption{Panel A: An $(a,sin(i))$ projection of the time evolution of fast 
clones or real V-type candidates in the Astraea region.  Panel B: the same for
clones of asteroids in the Merxia region. See caption of 
Fig.~\ref{fig: hansa_region} for a description of the used symbols. }
\label{fig:  Astraea_merxia_regions}
\end{figure*}

Fig.~\ref{fig:  Astraea_merxia_regions} displays the $(a,sin(i))$ 
projection of the time evolution of fast 
clones or real V-type candidates in the Astraea region (panel A) and
in the Merxia region (panel B).
One particle was able to cross the 3J:-1A mean-motion resonance and reached
the region of low-inclination V-type asteroids in the inner main belt
that Nesvorn\'{y} et al. (2008) showed to be unlikely to originate from 
the current Vesta family.   While this suggests that migration from the
inner main belt into the central is possible, it may also open
the possibility that the opposite happened, and that some of the 
low-inclination asteroids in the inner main belt may have originated
from the central main belt, possibly from the parent body of
the actual Merxia and Agnia families.

This seems to be supported by the fact that, once again, particles
with 100 m diameter appear to be able to cross the 11J:-4A and 8J:-3A 
mean-motion resonances relatively unharmed.  Communication from the Astraea
region to the Merxia and vice-versa seems therefore, at least a possibility.

Finally, we also observed minor changes of inclination caused by 
secular resonances for some particles, but not enough to reach the
Eunomia and Gefion regions.  The possibility of migrations between
the $sin(i) < 0.12$ and $0.12 < sin(i) < 0.3$ areas will be further 
investigated in the next subsection.

\subsection{Results of the ``slow simulations''}
\label{sec: slow_sims}

To further investigate the dynamical evolution of V-type candidates 
in the area, we performed ``long-term'' simulations of clones of the
same particles studied in Sect.~\ref{sec: fast_sims}.
These particles had the same parameters as in Sect.~\ref{sec: fast_sims}, 
except for the diameters and spin obliquities.
We took the WISE values of the diameters for the test particles 
for which such information was available, and we used
the methods described in Sect.~\ref{sec: V-type_groups} for the other
particles.  We also took six values of spin obliquities, $0^{\circ},
30^{\circ}, 60^{\circ}, 120^{\circ}, 150^{\circ}$ , and $180^{\circ}$, in 
order to sample different speeds of drifts caused by the Yarkovsky
force.  We integrated our test particles over 1 Byr, under the same
integration scheme used in Sect.~\ref{sec: fast_sims}.  Because
of the longer integration time used in these runs, we computed
proper elements every 4.9 Myr instead of the 1.2 Myr used
in Sect.~\ref{sec: fast_sims}.

In the Hansa region, contrary to the case of the fast simulations,
no particles managed to cross the 3J:-1A and 5J:-2A mean-motion resonances,
nor any particles was able to cross the dynamical barrier of the ${\nu}_6$
secular resonance.  All particles that reached the 8J:-3A mean-motion 
resonance were able to cross over, and minor changes in sine of inclination,
of up to 0.03, were observed after the passage of the 1J:-3S:1A three-body
resonance.  This suggests that the isolated V-type candidate in the Gallia
family region could be explained by migration from the Hansa family region
along with interaction with either the 8J:-3A or the 1J:-3S:1A mean-motion 
resonances.  Possible source mechanisms for the Hansa family region V-type
photometric candidates are still needed to explain the observed population
of objects.

Similar results were observed for asteroids in the Eunomia, Gefion, and Merxia
regions:  no particle survived the crossing of the the 3J:-1A, 5J:-2A 
mean-motion resonances, and no particle managed to pass the ${\nu}_6$ 
secular resonance barrier.  At least 80\% of the asteroids that 
reached the 8J:-3A and the 1J:-3S:1A mean-motion 
resonance were able to cross over.  Changes in inclination of up to
0.3 in $\sin{i}$ caused by interactions with 
the $2{\nu}_6-{\nu}_5$  secular resonances were also observed
in the Eunomia region, but this was not enough to allow particles to change
region (as changes occurred for values of increasing $i$, i.e., 
in the opposite direction to reach the Astraea region).  We did not 
observe any particle changing regions in inclination during the length of 
our simulations.

\begin{figure}

  \centering
  \centering \includegraphics [width=0.45\textwidth]{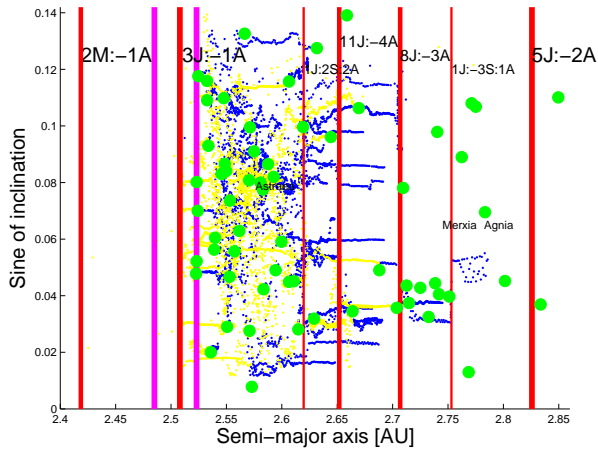}

\caption{An $(a,sin(i))$ projection of the time evolution of slow 
clones or real V-type candidates in the Astraea region.
See caption of Fig.~\ref{fig: hansa_region} for a 
description of the used symbols.}
\label{fig: Astraea_slow}
\end{figure}

Results were more interesting in the Astraea region. As discussed
in Sect.~\ref{sec: dyn_maps}, particles in this region
strongly interacted with the $2{\nu}_6-{\nu}_5$ non-linear
secular resonance, and this results in large oscillations of $sin(i)$ 
values.  Several particles are attracted by this powerful
resonance, creating a ``convergence zone'' of V-type
photometric candidates, roughly located between the 3J:-1A 
and 11J:-4A mean-motion 
resonances in semi-major axis (see also Fig.~\ref{fig: map_ai-cen} 
and the yellow dots associated with the same resonance for a better
definition of this region), and also shown in 
Fig.~\ref{fig: Astraea_slow},
that displays the time evolution in the $(a,sin(i))$ plane 
of clones of V-type candidates with obliquities of 
$0^{\circ}$ and $180^{\circ}$.  Yet, oscillations are
limited to the Astraea region. Only one particle managed to evolve
in the $2{\nu}_6-{\nu}_5$ to higher values of inclinations, but
it was soon lost because of its interaction with the 3J:-1A mean-motion
resonance.  No particle managed to cross the 3J:-1A and survive in the inner
main belt for more than 10 Myr.

Overall, our simulations seem to confirm our hypothesis of a bimodal source
for V-type asteroids in the central main belt, with the parent bodies
of the Eunomia family and of the Merxia and Agnia ones as possible sources.
The fact that no particle, during our simulations, switched in 
a stable fashion among inclination regions does not exclude that 
such orbital evolution is possible, but suggests, in our opinion, that 
these should be rare event.  To investigate
if dynamical mobility caused by other mechanisms such as close
encounters with massive asteroids could account for such dynamics, we
will present results of new simulations that also account for
this mechanism in the next section.

\section{Effects of close encounters with massive asteroids}
\label{sec: close_encounters}

In order to investigate the effects of close encounters with massive asteroids
we integrated all V-type photometric candidates over the gravitational
influence of all planets plus Ceres, Eunomia, and Juno, the most massive 
asteroids in the central main belt and some of the most effective perturbers
(see Carruba et al. (2013a) for a list of the first 40 
most massive asteroids in the main belt),
over 200 Myr.  We did not include the effect of 
Pallas, since Carruba et al. (2013a) showed that the long-term
effect of close encounters with this asteroid is less significant
than that of other, less inclined bodies, such as Ceres or Hygiea, and we
also did not simulated the Yarkovsky effect.  Since our goal was to
obtain a complete statistics of change in orbital elements due to close
encounters, and since this statistics does not strongly depend on the
model used (Carruba et al. 2012), we simply used SWIFT-SKEEL, the integrator
of the SyMBA package able to simulate close encounters among massive planets
and massless particles (Levison and Duncan 1994), without attempting to
include non-gravitational effects, and with the simulation set-up
discussed in Carruba et al. (2013a).

In this work we were mostly interested in dynamical mobility in inclination,
because these are the changes that can cause an asteroid to switch zones, so
``contaminating'' one source of basaltic material with another. 
Obtaining changes in proper inclination caused by close encounters only is
not a straightforward task.   That could be accomplished by 
obtaining proper elements before and after the encounter, and by trying to
isolate contributions to the changes in proper $i$ other then the close 
encounter (passage through mean-motion resonances, secular effects, etc.).
As a first step in our analysis, we just computed changes in osculating
$\sin(i)$, in order to have a preliminary estimate of the entity of
possible changes in proper $i$.

\begin{figure}

  \centering
  \centering \includegraphics [width=0.45\textwidth]{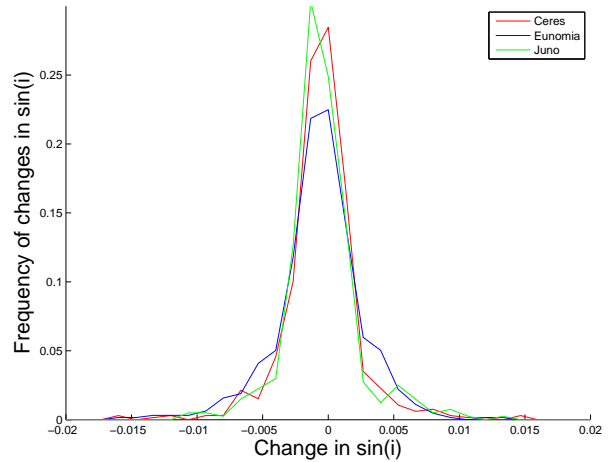}

\caption{Frequencies of changes in osculating
$\sin(i)$ caused by close encounters with (1) Ceres (red line), 
(15) Eunomia (blue line), and (3) Juno (green line).} 
\label{fig: di_changes}
\end{figure}

Our results are shown in Fig.~\ref{fig: di_changes}, where the red line
identifies the frequency distribution of changes in $\sin(i)$ caused by 
encounters with (1) Ceres, the blue line those caused by (15) Eunomia, 
and the green line those caused by (3) Juno.  During our simulations
we registered 657 encounters with Ceres, 636 with Eunomia, and 402  with
Juno, which, according to Carruba et al. (2013a), should be enough 
to obtain a statistics complete to a $1\sigma$ level, sufficient
to start an analysis of the long-term effect of close encounters.
As expected, most of the encounters cause a limited change in $\sin(i)$,
with less than 1\% of the encounters causing larger variations (the percentual
is higher for Ceres being this the most massive body).  Larger variations,
corresponding to encounters at very close relative distances or velocities,
are possible, but are very rare events.  Also, in order to allow for an
asteroid to switch zones in inclination, they should happen repeatedly and
in the ``right'' direction (to lower inclinations for Eunomia region
asteroids and to higher for Astraea region ones).  

In order to have a preliminary quantitative estimate of the long 
term effect of close encounters with massive asteroids we assume in 
first approximation that such mechanism could be treated as a random walk.
Under the assumption, as it seems to be the case for our observed distributions 
of changes in $sin(i)$, that the mean value of changes
is zero, there will be a $1\sigma$ (= 68.27\%) probability that  
the root mean square translation distance (or quadratic mean) after $n$ 
steps will fall between $\pm \sigma_{sin(i)} \sqrt{n}$, where $\sigma_{sin(i)}$
is the standard deviation of changes in $\sin(i)$, equal to 0.0029 for
the distribution of changes in $i$ caused by (1) Ceres, by far the main
perturber in the central main belt (Carruba et al. 2003, 
2013a).  Since Bottke et al. (2006) estimated
that the parent bodies of differentiated objects arrived
in the main belt no earlier than 4 Byr ago, using the value of 
$\sigma_{sin(i)}$ obtained for changes in $\sin(i)$ caused by (1) Ceres
over 200 Myr, and $n =20$, we obtained a root mean square translation
of 0.0130.  Considering our boundary between inclination regions at 
$\sin(i) = 0.12$, only 9 objects (7 in the Merxia/Agnia and 2 in the Eunomia
region) are in the region $0.12 \pm \sigma_{sin(i)} \sqrt{n}$ and 
could therefore experience a change in $sin(i)$ large enough to
allow for a change of region over 4 Byr, at $1\sigma$ level of probability, 
i.e., 7\% of the total.

Close encounters with massive asteroids may have allowed for some mixing
between asteroids from the Eunomia and the Merxia/Agnia regions, but 
a single source mechanism seems to be statistically unlikely.

%%%%%%%%%%%%%%%%%%%%%%%%%%%%%%%%%%%%%%%%%%%%%%%%%%%%%%%%%%%%
\section{Conclusions}
\label{sec: concl}

In this work we:

\begin{itemize}
\item Revised the current knowledge on V-type photometric candidates
(Carvano et al. 2010) and obtained new photometric candidates
with the approach of DeMeo and Carry (2013).  Overall, we identified
127 V-type photometric candidates in the central main belt, six of 
which belong to the Eunomia family, and four of which are Hansa family 
members.

\item Obtained distances with respect to (15) Eunomia and (808) Merxia
for all V-type photometric candidates, using the distance metrics 
of Zappal\'{a} et al. (1995). Three regions in inclination appear 
using this approach, suggesting a three-source model for the origin
of V-type asteroids in the central main belt:  highly-inclined
asteroids above the ${\nu}_6$ secular resonance (Hansa region),
asteroids with $0.12 < sin(i) < 0.3$ (Eunomia region), and asteroids
with $sin(i) < 0.12$ (Astraea region).  Additional subdivision can be
set by considering the presence of the 8J:-3A mean-motion resonance
(Gefion region in the Eunomia region, and Merxia region in the Astraea 
region).

\item Obtained a synthetic proper element 
dynamical map of the central main belt in the proper 
$(a,sin(i))$ plane for 19800 test particles, and studied
mean-motion and secular resonances (up to order six) in the
region. As hypothesized in Carruba et al. (2007a), the 
${\nu}_5-{\nu}+{\nu}_{16}$ could be a viable mechanism 
to reduce asteroids inclinations, but only for asteroids
in the Eunomia region, and not to values of 
$i$ in the Astraea region range.  
15 V-type photometric candidates 
are currently in librating 
states of the $2{\nu}_6-{\nu}_5$ secular resonance, 
and the local 
increase in number of these bodies in the Astraea 
region could therefore be an artifact of the local dynamics.

\item Studied the Yarkovsky-induced dynamical evolution of
clones of all V-type photometric candidates.  Our simulations
suggest that the current orbital distribution of these objects
could be explained with a two sources mechanism originating
in the parent bodies of the Eunomia and Agnia/Merxia families
(with a third possible source in the parent body of the Hansa
family).  With one (unstable) exception, no particle during 
our simulation managed to switch among inclination regions, and no 
particles with diameters larger than 1 km succeded in 
crossing the 3J:-1A mean-motion resonance
barrier, suggesting that asteroids from the Vesta family 
should be rare among central V-type photometric candidates.

\item Studied the effects of close encounters with 
(1) Ceres, (15) Eunomia, and (3) Juno for the V-type photometric
candidates.  Only less than 1\% of the studied encounters
caused a change in osculating $\sin(i)$ larger than 0.01 in module,
and no particle changed inclination zones over the length of
our integrations.  We estimated that, at $1\sigma$ level probability,
at most 7.0\% of the known V-type photometric candidates could have
changed inclination region over 4 Byr. 
  
\end{itemize}

%DISCUSS THAT IT IS NOT CLEAR IF THE DIFFERENTIATED OBJECTS ORIGINATED
%LOCALLY OR AT A < 2.0 AU (SCATTERED PLANETESIMALS, BOTTKE ET AL. 2006 NEED
%REF.).

Many new basaltic candidates have been identified since the three
originally suggested in 2007 (see also Carruba et al 2007a for a review
on V-type asteroids in the central main belt known at that time), and 
this allowed us to perform a more in depth analysis
of their possible origin.  Many more other asteroids may be 
missing or unobservable either because comminuted to sizes
below observational limits (Burbine et al. 1996), or because
of space weathering effects that obscured the basaltic signature
in their spectra (Wethrerill and Chapman, 1988).

For the currently known V-type candidates, we showed that a 
two source scenario, the parent bodies of the Eunomia plus
Merxia and Agnia families, could explain the origin of most of the basaltic
material in the central main belt, with a fraction that may come
from other sources (either from the inner main belt, diffusing
accross the 3J:-1A mean-motion resonance (Roig et
al. 2008), or because of some unusual mechanism of scattering, 
such very close encounters with massive asteroids, able to increase
the inclination of an object above the values of the center
of the ${\nu}_6$ secular resonance~\footnote{Highly inclined
V-type photometric candidates could also be explained
under the assumption of a third local source of basaltic
material, if we assume that the parent body of the Hansa
family was a differentiated or partially differentiated asteroid.}).
According to Bottke et al. (2006), the parent bodies of V-type asteroids
in the central main belt could possibly have been formed much closer than
their current orbital location, in the terrestrial planet region, and then 
have been scattered by close encounters with planetesimals in the early
phase of the Solar System formation, 4.0 Gyr ago or more.  Considering
that the number of scattered differentiated bodies is model dependent and 
ranges from one to several hundreds in Bottke et al. (2006) theory, depending
on the region of formation of differentiated bodies and on the minimum
value of diameter needed to insure differentiation, we believe
that a scenario in which two or three differentiated bodies were
injected into the central main belt and then disrupted by collisions
could explain the current observed distribution of V-type photometric
candidates in the central main belt~\footnote{Alternatively, depending
on the parameters of Bottke's theory, one could also imagine
that all central main belt V-type candidates were scattered from
the terrestrial planet region.}.

In this framework, it is of utmost importance to conduct 
a mineralogical analysis of the V-type candidates studied in
this work.  Sunshine et al. (2004) showed that Vestoids have much deeper band
depths than the HCP-rich S-type asteroids.  This could 
be used to discriminate the asteroids in the Astraea region
that originated from Vesta from those that originated
from the parent body of the Agnia and Merxia families, and could
be also important in assessing the origin of the highly inclined
V-type objects whose origin cannot be easily explained with scenarios
involving central main belt sources.  Overall, a lot of 
work remains to be done to understand the origin and the dynamical 
evolution of V-type asteroids in the central main belt. 

%%%%%%%%%%%%%%%%%%%%%%%%%%%%%%%%%%%%%%%%%%%%%%%%%%%%%%%%%%%%
\section*{Acknowledgments}
\label{sec: ack}

We are grateful to the reviewer of this article, Nicholas Moskowitz,
for suggestions and comments that significantly improved
the quality of this work.
We would like to thank the S\~{a}o Paulo State Science Foundation 
(FAPESP) that supported this work via the grant 11/19863-3, and the
Brazilian National Research Council (CNPq, grant 305453/2011-4).
This publication makes use of data products from the Wide-field 
Infrared Survey Explorer, which is a joint project of the University 
of California, Los Angeles, and the Jet Propulsion Laboratory/California 
Institute of Technology, funded by the National Aeronautics and Space 
Administration.  This publication also makes use of data products 
from NEOWISE, which is a project of the Jet Propulsion 
Laboratory/California Institute of Technology, funded by the Planetary 
Science Division of the National Aeronautics and Space Administration.

\section{Appendix}
\label{sec: app}

Table~\ref{table: V_carvano} reports the asteroid identification, 
its proper $a,e,$ and $sin(i)$, its absolute magnitude ($H$), 
diameter ($D$), and geometric
albedo ($p_V$), according to the WISE mission, when available, 
for all asteroids in the central main belt identified in Carvano
et al. (2010).   Asteroids without WISE data on $D$ and $p_V$ have been
assigned the mean value of geometric albedo of V-type
asteroids in the central main belt, and diameters computed according 
to Eq.~\ref{eq: D_H_pv} (for simplicity, 
we do not report these data in the Table).  
For asteroids likely to be in a secular resonance configuration,
we also report the resonance name.
Finally, we specified the region to which
the asteroid belongs to, according to the classification scheme of
Sect.~\ref{sec: taxonomy-sdss}, and listed asteroids accordingly.

\onecolumn 
\begin{center}
\begin{longtable}{cccccccc}
\caption{List of V-type photometric candidates in the central Main-Belt} \\
\hline
                  &             &         &             &   &        &       & 
\\
Asteroid id.      &	$a$	& $e$     & $\sin(i)$	& H & D (km) & $p_v$ & 
Seculare resonance name\\
                  &             &         &             &   &        &       &  \\
\hline
\endfirsthead
\multicolumn{7}{c}%
{\tablename\ \thetable\ -- \textit{Continued from previous page}} \\
\hline
                  &             &         &             &   &        &      &  
\\
Asteroid id.      &	$a$	& $e$     & $\sin(i)$	& H & D (km) & $p_v$ & 
Secular resonance name\\
                  &             &         &             &   &        &      &
\\
\hline
\endhead
\hline \multicolumn{7}{r}{\textit{Continued on next page}} \\
\endfoot
\hline
\endlastfoot
\label{table: V_carvano}
\textbf{Hansa region} \\
\hline
  44257 & 2.5472 & 0.0392 & 0.3822 & 13.70 & 4.6450 & 0.2710  & \\
  62002 & 2.7586 & 0.2264 & 0.4356 & 14.90 &        &         &\\
  87908 & 2.6833 & 0.0552 & 0.3718 & 14.10 & 3.8760 & 0.2456  &\\
 118748 & 2.6130 & 0.0763 & 0.3699 & 16.90 &        &         &\\
 122662 & 2.5774 & 0.0674 & 0.3717 & 15.00 & 3.0520 & 0.2741  &\\
 182339 & 2.6788 & 0.0338 & 0.3772 & 15.79 &        &         &\\
 217433 & 2.7743 & 0.0840 & 0.3071 & 14.97 &        &         & $3{\nu}_6-{\nu}_5$\\
\hline
 \textbf{Eunomia region} \\
\hline
 21238 & 2.5411 & 0.1371 & 0.1866 & 12.90 & 5.2210 & 0.3729  &\\
 24014 & 2.5609 & 0.1709 & 0.2241 & 13.10 & 6.5670 & 0.2357  &\\
 40521 & 2.5311 & 0.0458 & 0.2159 & 14.90 & 2.6370 & 0.2785  &\\
 44447 & 2.5701 & 0.1641 & 0.2358 & 14.70 & 3.7360 & 0.1668  & $3{\nu}_6-{\nu}_5$\\
 55550 & 2.5457 & 0.1444 & 0.2298 & 15.30 &        &         &\\
 56904 & 2.6338 & 0.1733 & 0.2310 & 14.90 &        &         & $2{\nu}_6+{\nu}_{16} = z_2$\\
 81147 & 2.5326 & 0.0927 & 0.1755 & 15.00 &        &         & ${\nu}_6+{\nu}_{16} = z_1$\\
 84021 & 2.6790 & 0.0911 & 0.2539 & 15.40 &        &         & $3{\nu}_6-{\nu}_5$\\
 88533 & 2.6581 & 0.1374 & 0.2021 & 15.10 &        &         &\\
 90649 & 2.6587 & 0.1606 & 0.1391 & 15.80 &        &         &\\
 93620 & 2.6327 & 0.1437 & 0.2460 & 14.70 & 2.5870 & 0.3173  & $3{\nu}_6-{\nu}_5$, $2{\nu}_6+{\nu}_{16} = z_2$ \\
%93981 & 2.6625 & 0.0192 & 0.2263 & 13.85 & 3.3760 & 0.4268  &\\
108139 & 2.5374 & 0.2553 & 0.2150 & 14.90 & 3.3020 & 0.1777  &\\
110998 & 2.6162 & 0.0392 & 0.2687 & 15.00 &        &         &\\
120236 & 2.6086 & 0.1994 & 0.2630 & 16.90 &        &         &\\
126729 & 2.6389 & 0.1217 & 0.2566 & 15.60 & 2.1150 & 0.2272  &\\
147640 & 2.5686 & 0.1568 & 0.2264 & 15.30 &        &         &\\
170007 & 2.6907 & 0.1364 & 0.1640 & 16.60 &        &         &\\
170015 & 2.6871 & 0.2185 & 0.1529 & 15.80 &        &         & $2{\nu}_6-{\nu}_5$\\
189759 & 2.5743 & 0.0483 & 0.1589 & 15.47 &        &         &\\
196445 & 2.5418 & 0.1995 & 0.1579 & 16.15 &        &         &\\
196484 & 2.5665 & 0.1712 & 0.1326 & 16.27 &        &         &\\
201693 & 2.5669 & 0.0534 & 0.2401 & 15.69 &        &         &\\
202118 & 2.5364 & 0.0713 & 0.2359 & 16.14 &        &         &\\
203669 & 2.5422 & 0.1875 & 0.2583 & 15.51 & 1.7190 & 0.3441  &\\
208783 & 2.5395 & 0.2659 & 0.2236 & 15.60 & 2.4620 & 0.1676  &\\
238208 & 2.7029 & 0.0674 & 0.1584 & 16.15 & 2.2480 & 0.1269  &\\
239099 & 2.6036 & 0.0754 & 0.2069 & 16.19 &        &         &\\
242039 & 2.6864 & 0.0525 & 0.1818 & 15.69 &        &         &\\
243029 & 2.6860 & 0.0776 & 0.2521 & 15.67 & 2.4220 & 0.1580  &\\
249831 & 2.6400 & 0.1598 & 0.1715 & 15.84 &        &         &\\
262637 & 2.6259 & 0.2339 & 0.2291 & 16.20 &        &         &\\
271071 & 2.5631 & 0.0878 & 0.2594 & 15.74 &        &         &\\
275687 & 2.5668 & 0.1438 & 0.2269 & 16.67 &        &         &\\
279983 & 2.6813 & 0.0669 & 0.2394 & 16.24 &        &         &\\
302936 & 2.5122 & 0.1003 & 0.1881 & 16.63 &        &         &\\
318027 & 2.6694 & 0.2368 & 0.2289 & 16.54 &        &         &\\
324579 & 2.6805 & 0.0786 & 0.2369 & 16.60 &        &         &\\
341273 & 2.7835 & 0.0530 & 0.2346 & 15.74 &        &         &\\
363180 & 2.6317 & 0.1740 & 0.1275 & 16.78 &        &         &\\
363846 & 2.6225 & 0.1295 & 0.2305 & 16.06 &        &         &\\
367188 & 2.6902 & 0.0257 & 0.2079 & 15.90 &        &         &\\
\hline
 \textbf{Astraea region} \\
\hline
22893 & 2.5326 & 0.1893 & 0.1091 & 15.10 &         &         &\\
29720 & 2.6638 & 0.0846 & 0.0345 & 15.30 &         &         & ${\nu}_6+{\nu}_{16} = z_1$\\
46626 & 2.5359 & 0.1602 & 0.0200 & 14.60 &         &         &\\
57439 & 2.5617 & 0.0677 & 0.0629 & 15.40 &         &         &\\
58424 & 2.5337 & 0.1850 & 0.0930 & 15.50 &         &         &\\
67299 & 2.5751 & 0.2692 & 0.0910 & 14.90 &         &         &\\
68635 & 2.5808 & 0.2461 & 0.0800 & 16.30 & 2.0110 & 0.3633   &\\
89387 & 2.5507 & 0.1417 & 0.0290 & 16.00 &         &         &\\
93580 & 2.6443 & 0.2865 & 0.0961 & 16.70 &         &         &\\
102973 & 2.6108 & 0.1347 & 0.0451 & 15.90 &        &         &\\
108901 & 2.5944 & 0.1206 & 0.0490 & 16.60 &        &         & $2{\nu}_6-{\nu}_5$, ${\nu}_6+{\nu}_{16} = z_1$ \\
119360 & 2.5498 & 0.1403 & 0.0841 & 16.60 &        &         & $2{\nu}_6-{\nu}_5$\\
121046 & 2.5226 & 0.1289 & 0.0478 & 16.70 &        &         &\\
122758 & 2.5705 & 0.1892 & 0.0808 & 17.20 &        &         & $2{\nu}_6-{\nu}_5$\\
123441 & 2.6291 & 0.0710 & 0.0319 & 16.30 &        &         &\\
127242 & 2.5398 & 0.1603 & 0.0605 & 15.70 & 2.0490 & 0.2014  &\\
129632 & 2.5325 & 0.2441 & 0.1159 & 16.20 & 1.5120 & 0.3373  &\\
129633 & 2.5231 & 0.1693 & 0.0802 & 16.30 &        &         &\\
134783 & 2.6065 & 0.0105 & 0.0449 & 16.30 &        &         & $2{\nu}_6-{\nu}_5$\\
136823 & 2.6192 & 0.2200 & 0.0996 & 16.20 &        &         & $2{\nu}_6-{\nu}_5$\\
137139 & 2.5456 & 0.2783 & 0.0830 & 16.20 &        &         &\\
138114 & 2.5476 & 0.2404 & 0.1098 & 16.80 &        &         &\\
148699 & 2.6695 & 0.1238 & 0.1063 & 16.10 &        &         &\\
149588 & 2.5390 & 0.0699 & 0.0563 & 16.50 &        &         &\\
152062 & 2.5713 & 0.2847 & 0.0995 & 16.40 &        &         &\\
154704 & 2.5528 & 0.1659 & 0.0467 & 16.50 &        &         &\\
158290 & 2.5709 & 0.1111 & 0.0276 & 17.00 &        &         & $2{\nu}_6-{\nu}_5$\\
160623 & 2.6065 & 0.1961 & 0.1157 & 15.80 &        &         &\\
161465 & 2.7037 & 0.0624 & 0.0357 & 15.80 &        &         &\\
177717 & 2.5485 & 0.0615 & 0.0866 & 15.91 &        &         &\\
181565 & 2.6149 & 0.1103 & 0.0281 & 16.40 &        &         &\\
187458 & 2.5927 & 0.1854 & 0.0820 & 15.14 &        &         & $2{\nu}_6-{\nu}_5$\\
189410 & 2.5876 & 0.2541 & 0.0865 & 14.64 &        &         & $2{\nu}_6-{\nu}_5$\\
225791 & 2.5835 & 0.1124 & 0.0423 & 17.22 &        &         & $2{\nu}_6-{\nu}_5, 2{\nu}_6+{\nu}_{17}$\\
240369 & 2.5241 & 0.0551 & 0.0700 & 16.58 &        &         &\\
271802 & 2.6880 & 0.0836 & 0.0490 & 17.42 &        &         &\\
287090 & 2.5247 & 0.2083 & 0.1176 & 17.68 &        &         &\\
293407 & 2.5829 & 0.1698 & 0.0775 & 17.42 &        &         & $2{\nu}_6-{\nu}_5$\\
327872 & 2.5996 & 0.1696 & 0.0591 & 17.17 &        &         &\\
329973 & 2.5729 & 0.0895 & 0.0078 & 17.57 &        &         & $2{\nu}_6-{\nu}_5$\\
%348518 & 2.7419 & 0.0538 & 0.0405 & 16.86 & 1.1767 & 0.2300  &\\
366838 & 2.5532 & 0.2321 & 0.0736 & 18.11 &        &         &\\
370092 & 2.5231 & 0.2077 & 0.0522 & 17.74 &        &         &\\
370122 & 2.5573 & 0.1022 & 0.0557 & 17.44 &        &         &\\
\hline
 \textbf{Gefion Region}\\
\hline
   7302 & 2.8074 & 0.1468 & 0.1595 & 12.10 &10.6290& 0.2082  &\\
  22308 & 2.7734 & 0.1126 & 0.2119 & 14.00 & 5.8230 & 0.1573  &\\
  37325 & 2.7398 & 0.1831 & 0.1670 & 14.30 &       &         &\\
  66905 & 2.7462 & 0.1458 & 0.2291 & 15.30 &       &         &\\
  87110 & 2.7112 & 0.1606 & 0.2390 & 15.00 &       &         &\\
  93322 & 2.7432 & 0.0990 & 0.2080 & 15.30 & 3.2540 & 0.1829 &\\
 104359 & 2.7714 & 0.1360 & 0.1562 & 15.10 &       &         &\\
 113194 & 2.7206 & 0.1037 & 0.1646 & 15.30 &       &         & $2{\nu}_6-{\nu}_5$\\
 119686 & 2.7150 & 0.1796 & 0.1632 & 15.30 &       &         & $2{\nu}_6-{\nu}_5$\\
 126955 & 2.7352 & 0.1396 & 0.1557 & 15.00 &       &         &\\
 133990 & 2.7751 & 0.0255 & 0.2538 & 15.50 &       &         &\\
 150512 & 2.7216 & 0.1042 & 0.2316 & 15.90 &       &         &\\
 179216 & 2.7529 & 0.0648 & 0.1884 & 15.61 &       &         &\\
 182385 & 2.7271 & 0.1809 & 0.1665 & 15.00 & 1.8520 & 0.2464 & $2{\nu}_6-{\nu}_5$\\
 186607 & 2.7570 & 0.0659 & 0.2424 & 15.25 &       &         &\\
 202242 & 2.7379 & 0.1596 & 0.2211 & 15.28 & 2.2360 & 0.2230 &\\
 206011 & 2.7240 & 0.0334 & 0.2325 & 15.42 & 2.4360 & 0.1713 & ${\nu}_5-{\nu}_6+{\nu}_{16}$\\
 233474 & 2.7835 & 0.1218 & 0.1596 & 15.91 &       &         &\\
 258311 & 2.7856 & 0.0858 & 0.1600 & 16.26 &       &         & $2{\nu}_6+{\nu}_{17}$ \\
 340344 & 2.7405 & 0.1311 & 0.1529 & 16.99 &       &         &\\
 \hline
 \textbf{Merxia region} \\
\hline
 36118 & 2.7095 & 0.0742 & 0.0781 & 13.40 & 4.9980 & 0.3385  &\\
 36590 & 2.8336 & 0.0473 & 0.0369 & 15.40 & 2.7020 & 0.1674  &\\
 48558 & 2.7624 & 0.1386 & 0.0890 & 15.40 &        &         &\\
 85402 & 2.7325 & 0.0577 & 0.0325 & 15.00 &        &         &\\
 88727 & 2.8015 & 0.1192 & 0.0452 & 15.00 &        &         &\\
 88754 & 2.7384 & 0.0744 & 0.0444 & 14.90 &        &         & ${\nu}_6+{\nu}_{16} = z_1$\\
 95536 & 2.7512 & 0.1011 & 0.0397 & 15.90 &        &         &\\
155236 & 2.7148 & 0.0620 & 0.0374 & 16.20 &        &         &\\
171403 & 2.7834 & 0.0504 & 0.0695 & 15.80 &        &         &\\
126639 & 2.7250 & 0.0401 & 0.0428 & 15.80 &        &         &\\
208899 & 2.7751 & 0.0392 & 0.1068 & 15.73 &        &         &\\
209318 & 2.7403 & 0.0419 & 0.0979 & 16.78 &        &         &\\
233503 & 2.7126 & 0.0323 & 0.0436 & 16.06 &        &         &\\
243256 & 2.7713 & 0.0901 & 0.1080 & 16.05 & 3.5200 & 0.0518  &\\
289970 & 2.7687 & 0.0494 & 0.0130 & 16.86 &        &         &\\
348518 & 2.7419 & 0.0538 & 0.0405 & 16.86 &        &         &\\
\bottomrule%\hline
\end{longtable}
\end{center}

\twocolumn

\bsp

\label{lastpage}

\end{document}